\documentclass[journal]{IEEEtran}
\usepackage{bookmark}
\usepackage{tabularx}
\usepackage{lineno}
\usepackage{chngpage}
\usepackage{float}
\usepackage{graphicx}
\usepackage{epsfig}
\usepackage{epstopdf}
\usepackage{array}
\usepackage{diagbox}
\usepackage{epstopdf}
\usepackage{algorithm}
\usepackage{algorithmicx}
\usepackage{algpseudocode}
\usepackage{threeparttable}
\usepackage{dsfont}
\usepackage{subfigure}
\usepackage{multirow}
\usepackage{longtable}
\usepackage[table,xcdraw]{xcolor}
\usepackage{url}
\usepackage{booktabs}
\usepackage{multirow}
\usepackage{algpseudocode}
\usepackage{amsmath}
\usepackage{amssymb}
\usepackage{booktabs}
\usepackage{amsthm, amssymb}
\theoremstyle{definition}
\newtheorem{definition}{Definition}

\ifCLASSINFOpdf
\else
\fi
\begin{document}
%
\title{Matching Algorithms: Fundamentals, Applications and Challenges}
%
%
%

\author{Jing Ren, Feng Xia,~\IEEEmembership{Senior Member,~IEEE,}
        Xiangtai Chen, Jiaying Liu, Mingliang Hou, \\Ahsan Shehzad, Nargiz Sultanova, and Xiangjie Kong,~\IEEEmembership{Senior Member,~IEEE}
\thanks{This work is partially supported by National Natural Science Foundation of China under Grant No. 61872054. \textit{(Corresponding author: Feng Xia.)}}
\thanks{J. Ren, F. Xia and N. Sultanova are with the School of Engineering, IT and Physical Sciences, Federation University Australia, Ballarat, VIC 3353,
Australia (e-mail: ch.yum@outlook.com; f.xia@ieee.org; n.sultanova@federation.edu.au).}
\thanks{X. Chen, J. Liu, M. Hou and A. Shehzad are with the School of Software, Dalian University of Technology, Dalian 116620, China (e-mail: chenxiangtai@outlook.com; jiaying\_liu@outlook.com; teemohold@outlook.com; ahsan.shehzad@outlook.com).}

\thanks{X. Kong is with College of Computer Science and Technology, Zhejiang University of Technology, Hangzhou 310023, China (e-mail: xjkong@ieee.org).}
\thanks{\textcopyright ~2021 IEEE.  Personal use of this material is permitted.  Permission from IEEE must be obtained for all other uses, in any current or future media, including reprinting/republishing this material for advertising or promotional purposes, creating new collective works, for resale or redistribution to servers or lists, or reuse of any copyrighted component of this work in other works.}
}

%
%

\markboth{IEEE TRANSACTIONS ON EMERGING TOPICS IN COMPUTATIONAL INTELLIGENCE}%
{Ren \MakeLowercase{\textit{et al.}}: Matching Algorithms: Fundamentals, Applications and Challenges}
%



\maketitle

\begin{abstract}
Matching plays a vital role in the rational allocation of resources in many areas, ranging from market operation to people's daily lives. In economics, the term matching theory is coined for pairing two agents in a specific market to reach a stable or optimal state. In computer science, all branches of matching problems have emerged, such as the question-answer matching in information retrieval, user-item matching in a recommender system, and entity-relation matching in the knowledge graph. A preference list is the core element during a matching process, which can either be obtained directly from the agents or generated indirectly by prediction. Based on the preference list access, matching problems are divided into two categories, i.e., \emph{explicit matching} and \emph{implicit matching}. In this paper, we first introduce the matching theory's basic models and algorithms in explicit matching. The existing methods for coping with various matching problems in implicit matching are reviewed, such as retrieval matching, user-item matching,  entity-relation matching, and image matching. Furthermore, we look into representative applications in these areas, including marriage and labor markets in explicit matching and several similarity-based matching problems in implicit matching. Finally, this survey paper concludes with a discussion of open issues and promising future directions in the field of matching.
\end{abstract}


%
\IEEEpeerreviewmaketitle

\section{Introduction}

With the rapid growth of science and technology, an extensive repository of matching demands have emerged in different fields. Therefore, identifying, analyzing, and managing resources have become increasingly challenging. The term "matching" is generally defined as two objects suitably paired together or having the same appearance. Inspired by this definition, this paper's matching problem is defined as looking for a method to pair two or more objects together so that the pairs are suitably matched or have a similar appearance.
Traditional matching problems (classified as explicit matching in this paper) emphasize finding the most suitable object according to their preference list. The final goal is to reach a stable or optimal state in a specific market~\cite{gale1962college}. This kind of problem is mainly studied in the fields of economics and mathematics. However, in the age of big data, obtaining everyone's preference list is time-consuming and nearly impossible. Therefore, matching problems nowadays in computer science pay more attention to predicting users' preference list.

Unlike the commodity market, money is not involved during the process in the matching markets. In a broad sense, matching can be found in many disciplines and fields~\cite{wang2018learning,akbarpour2020thickness}. For example, face recognition is an application of image matching~\cite{zhao2003face}, and information retrieval needs the technology of matching based on text similarity~\cite{berger2017information}. For an explicit matching, the preference lists for the matching process are provided by the agents themselves, and the ultimate goal of an explicit matching problem is to reach a stable or optimal state in a specific context~\cite{hoessler2020stable,che2019stable}. As for an implicit matching, we need to calculate the matching score between every two agents, thereby obtaining the preference list by ranking their matching scores~\cite{xia2014mvcwalker,wang2020venue}.

Gale and Shapley~\cite{gale1962college} first proposed the concept of matching theory in 1962, which is a mathematical framework based on the game theory, and applied it to the marriage market and college admission. Since then, this theory has attracted the interests of a panoply of economists as it can solve many matching problems where the money is not involved~\cite{kazmi2020distributed,ismaili2019weighted}. When only one side of the agents has a preference list, it is regarded as one-sided matching, while matching with preference lists on both sides is a two-sided matching. The two-sided matching theory considers that two sets of matching agents are selfish and rational. They seek to be matched to each other, for example, men and women in a marriage market, hospitals, and interns in a labor market, or buyers and sellers in an auction market. Each agent has a preference list based on their preferences over the set of agents on the opposite side. In such matching markets, stability is considered a fundamental requirement of successful matching. This concept implies that no pair of agents attempt to leave their current partners and form a new pair with each other. The optimal matching problems, like D2D communications in the wireless network, are mainly defined as optimization problems whereby the optimization objective function and constraints are given in accordance with the context.

\begin{figure*}[htb]
	\centering
	\includegraphics[width=0.9\textwidth]{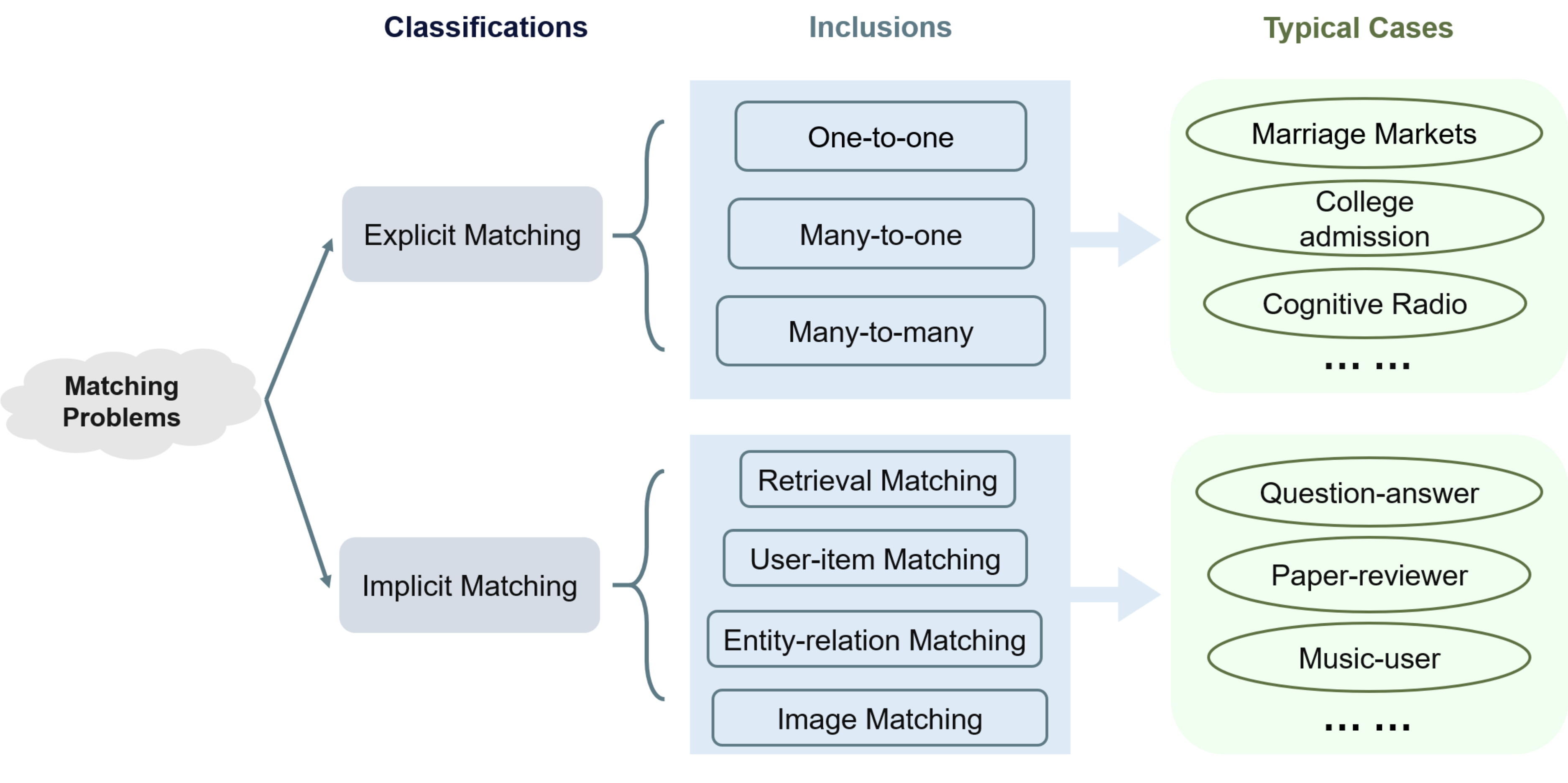}\\
	\caption{The classifications and applications of matching problems.}\label{framework.pdf}
\end{figure*}

In most cases, the preference lists of agents cannot be obtained directly due to the data sets' limitations. Therefore, the generation of preference lists becomes the first step of matching. In the field of information retrieval, a wide spectrum of methods have been proposed to calculate the similarity between texts~\cite{devlin2018bert}. This process is vital for many practical applications, such as question-answer matching in search engines. Another important field is recommender systems~\cite{asabere2014improving,xia2013mobile}, where we need to judge the preference of users, such as recommending music and movies to users, recommending merchandise to consumers, or recommending papers to reviewers. Furthermore, with the help of knowledge graph~\cite{liu2020web,liu2019two}, applications such as link prediction in relationship network~\cite{xia2014exploiting,bedru2020big} and user interest prediction in recommender system~\cite{xia2014socially,kong2017time} can be realized more accurately.

Instead of selecting only one research direction, either explicit matching problems~\cite{abdulkadiroglu2013matching, hakimov2019experiments} or implicit matching problems~\cite{ma2020image,gomaa2013survey}, this paper comprehensively reviews both kinds of matching problems in different fields. The advantage of this survey is to provide readers with more branches in both economics and computer science, such as stable matching theory, recommendation system, information retrieval, knowledge graph, and face alignment. Besides, readers who are not familiar with the stable matching theory may get a new idea of applying it to solve real-world problems in different fields. To the best of our knowledge, this paper is the first effort in providing a review and classification of matching problems according to the access of preference lists. In explicit matching problems such as marriage market and medical intern, the stable matching theory algorithm (e.g., deferred acceptance algorithm) and its variants are used to match selfish, rational agents with known preference lists. On the other hand, implicit matching problems aim to match entities by generating preference lists, like paper-reviewer matching, user-merchandise matching, and question-answer matching. Explicit matching is mainly introduced from three perspectives: one-to-one, many-to-one, and many-to-many matching. As for implicit matching, we only present some typical matching problems and their methods in computer science. The main details of the classification of matching problems are shown in Fig.~\ref{framework.pdf}.

The rest of this paper is structured as follows. Section~\ref{sec2} presents the concepts and methods of explicit matching, and Section~\ref{sec3} describes parts of emerging issues and methods in implicit matching. Comprehensive applications in both explicit and implicit matching are introduced in Section~\ref{sec4}. We discuss some future trends and challenges in Section~\ref{sec5}, and conclude this paper in Section~\ref{sec6}. We summarize all algorithms of explicit and implicit matching problems surveyed in this paper in Table~\ref{sum}. It should be noted that there is no systematic classification in explicit matching algorithms. Therefore, references of explicit matching shown in this table are classified according to real-world applications.

\begin{table*}[htb]
	\caption{Summarisation of explicit and implicit matching algorithms and applications}
	\begin{tabular}{|p{2cm}<{\centering}|p{3cm}<{\centering}|p{4.5cm}<{\centering}|p{3.9cm}<{\centering}|p{3.1cm}<{\centering}|}
		\hline
		\textbf{Category} & \textbf{Sub-category} & \textbf{Alogorithms} & \textbf{references} & \textbf{Applications}\\
		\hline
		\multirow{4}{*}{Explicit Matching}	&
		One-to-one Matching& -   & \cite{gale1962college, matching1992rothalvin, becker1973marriage, bergstrom1993courtship}& marriage market \\ \cline{2-5}
		
		&	\multirow{2}{*}{Many-to-one Matching} &\multirow{2}{*}{ - } & \cite{ che2019stable,ismaili2019weighted, kelso1982many,  rosen1981economics,roth1984stability}& job matching \\ \cline{4-5}
		&& & \cite{kremer1993ring,gabaix2008has,tervio2008difference} &pay matching\\ \cline{2-5}
		&	\multirow{2}{*}{Many-to-many Matching} & \multirow{2}{*}{ Optimization algorithms}  &\multirow{2}{*}{ \cite{hoessler2020stable,naparstek2014distributed,gu2014cheating,leshem2011multichannel} } & cognitive radio networks; D2D communications\\
		\hline
		\multirow{16}{*}{Implicit Matching}	&
		\multirow{4}{*}{Retrieval Matching}& Traditional matching algorithms & \cite{ramos2003using, salton1988term, robertson2009probabilistic, dumais2004latent, hofmann1999probabilistic, blei2003latent}& \\ \cline{3-4}
		&& Representation-based algorithms & \cite{huang2013learning, shen2014latent, palangi2014semantic, ChoiYL18} & machine translation; \\\cline{3-4}
		&& \multirow{2}{*}{Interaction-based algorithms} & \cite{hu2014convolutional, bengio2009learning, pang2016text, wan2016match, cho2014learning, parikh2016decomposable, wan2016deep, yang2016anmm, kim2019semantic, mitra2017learning} &expertise matching; question-answer matching   \\
		\cline{2-5}
		
		&\multirow{3}{*}{User-item Matching} &Basic algorithms& \cite{Asocft, tran2018regularizing}  & \\ \cline{3-4}
		&&Represenation-based algorithms&  \cite{Autorec, Cdaft, SVDpp, Dmfm, Acf} & recommendation systems \\ \cline{3-4}
		&&Matching function-based algorithms& \cite{Ncf, Lrmlv, Nfmfs, Afml} &  \\ \cline{2-5}
		&\multirow{3}{*}{Entity-relation Matching} & Factorization-based algorithms & \cite{BCTF, RESCAL, Tetac, ANALOGY} &recommendation systems;   \\\cline{3-4}
		&& Neural network-based algorithms &\cite{SME, NTN, MLP, NAM} & knowledge fusion; \\\cline{3-4}
		&& Translational distance based algorithms & \cite{TransE, TransH, TransR, TransD, TransA} &  information retrieval \\\cline{2-5}
		&	\multirow{3}{*}{Image Matching} & Area-based algorithms  & \cite{levine1973computer, gruen1985adaptive, wu2019fast, ishiyama2018fast, yang2018image} &  robot vision;\\ \cline{3-4}
		&& \multirow{2}{*}{Feature-based algorithms} & \cite{harris1988combined, lowe2004distinctive, bay2008speeded, rublee2011orb, babri2016feature, karami2017image, taigman2014deepface, schroff2015facenet, he2016deep, wang2019racial} & object recognition;
		medical image diagnosis \\
		\hline
		
	\end{tabular}

\end{table*}	\label{sum}

\section{Explicit Matching}~\label{sec2}
\begin{figure}[htb]
	\centering
	\includegraphics[width=0.45\textwidth]{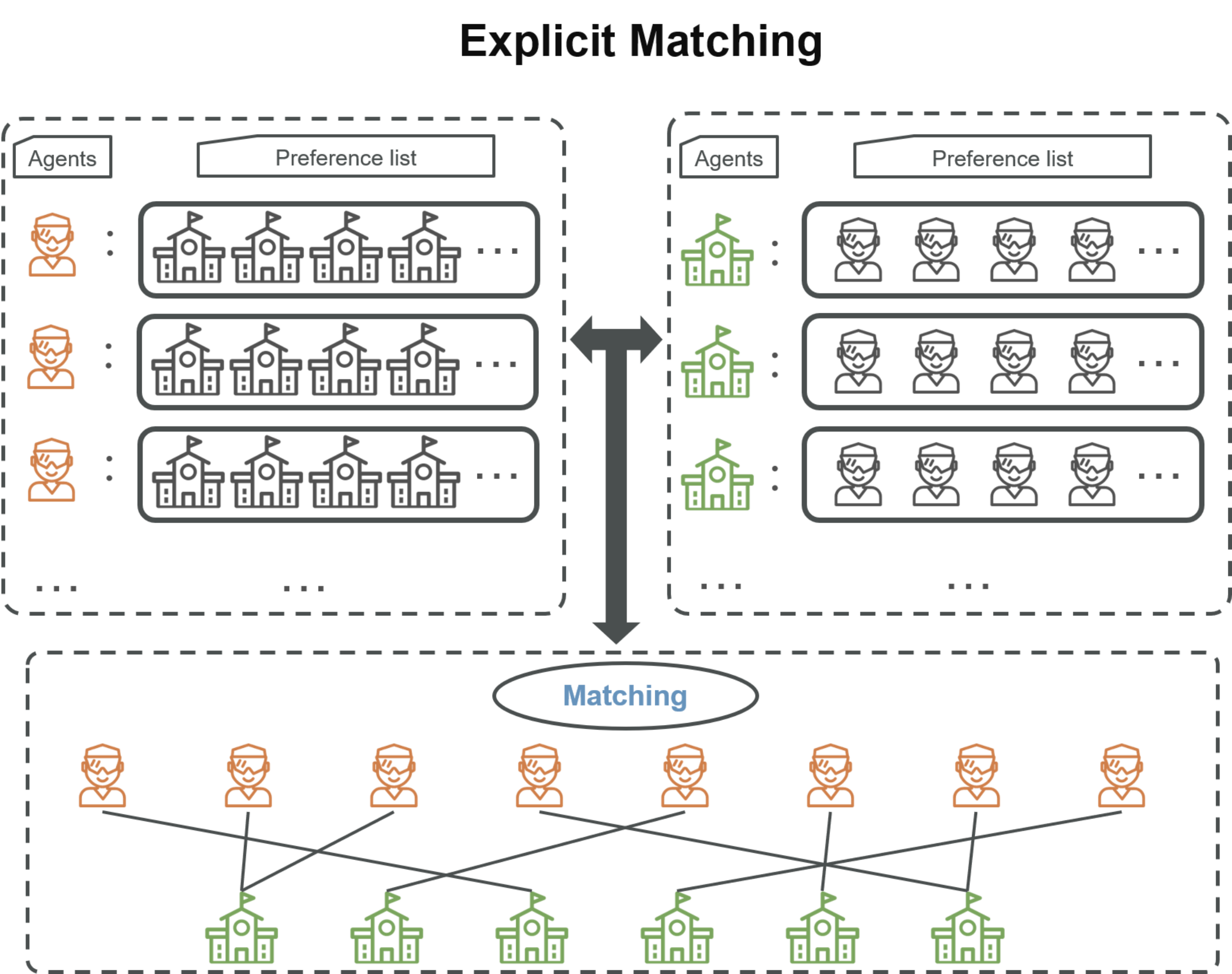}\\
	\caption{An illustration of student-school matching (Explicit Matching).}\label{explicitmatching}
\end{figure}

Explicit matching refers to problems in which the agents themselves give preference lists. This section provides the classification of these kinds of problems into three types; namely one-to-one, many-to-one, and many-to-many. Then, the process of the famous deferred acceptance algorithm is introduced in detail. Besides, an illustration of many-to-one matching is shown in Fig.~\ref{explicitmatching}.

\subsection{One-to-One}
We start by introducing the basic matching theory in the classical one-to-one marriage model~\cite{matching1992rothalvin}. There are two sets of agents in this model, men and women, represented by $M=\{m_1,m_2,...,m_n \}$ and $W=\{w_1,w_2,...,w_k\}$, respectively. Each agent has a complete preference list over the agents on the other side. The one-to-one matching of men and women is the outcome of the marriage problem. Assume that an agent $m$ has a preference list $Per(m)= w_1,w_2,m,w_3,w_4,...$. This means that they prefer agent $w_1$ to $w_2$ and prefer remaining single ($m$) over partnering with $w_3$ or $w_4$. We use $w_i >_m w_j$ to express that an agent $m$ prefers agent $w_i$ to $w_j$. In addition, $w_i \geq_m w_j$ means that for the agent $m$, the preference for $w_i$ is no less than for $w_j$.

The One-to-one matching model (taking marriage market as an example here) is defined as follows:

\begin{definition}
	\textit{
	An outcome of one-to-one marriage model is a matching $\mu$ from $M \cup W$ to $M \cup W$ such that:
	\begin{itemize}
		\item For any $m\in M$, $\mu(m)\in W\cup\{m\}$,
		\item For any $w\in W$, $\mu(w)\in M\cup\{w\}$,
		\item For any $m\in M$ and $w\in W$, $w=\mu\left(m\right)$ if and only if $m=\mu\left(w\right)$.
	\end{itemize}}
\end{definition}
Note that in the process of one-to-one matching problems, every man $m$ can only be matched with one woman $w$, and agent $m$ remains unmatched (single) if $\mu(m)=m$. The goal of an explicit matching is to reach stable status for all pairs:
\begin{definition}
	\textit{
		A matching $\mu$ is pairwise stable if any individuals or pairs do not block it.}
\end{definition}
\begin{definition}
	\textit{
	A matching $\mu(x)$ is blocked if $x$ prefers remaining single to being matched with someone (for any $x\in M\cup W$, $x>_x \mu(x)$). Both matchings $\mu(w)$ and $\mu(m)$ are blocked if $w$ and $m$ prefer each other to their current partners ($w>_m \mu(m)$ and $m>_w \mu(w)$).}
\end{definition}
The One-to-one matching is to obtain the most-preferred allocation that completely suits both agents' preference lists and reaches a stable status at the same time.

\subsection{Many-to-One}
Many-to-one or one-to-many matching indicates that agents of one side are allowed to be matched with more than one agent on the other side. Typical examples of many-to-one matching arise in the person-institution matching (shown in Fig.~\ref{explicitmatching}) problems such as student-college or doctor-hospital matching \cite{shimada2020multi}. The rule behind this kind of matching is that agents on one side (e.g., institutions) can provide many of the same positions for the agents on the other side (e.g., students), but the reverse is not valid.
\par Firstly, many-to-one matching model requires two finite disjoint sets, $P=\{p_1,p_2, ..., p_n\}$ and $V=\{v_1,v_2, ..., v_m\}$ representing the sets of people and institutions. Similar to one-to-one matching, $Per(v)= p_1,p_2,v,p_3,p_4,...$ means that institution $v$ prefers $p_1$ to $p_2$ and prefers keeping the position unfilled over other people like $p_3$ and $p_4$.
\par Unlike one-to-one matching, each institution has a positive quota $q$ to represent the maximum number of people it can be matched with. For a given institution $v$, its quota can be written as $q_v$. Therefore, in many-to-one matching, one person can only be matched with one institution and one institution can be matched with a fixed number of people ($q_v$) at most. Unmatched positions in the preferences of people or institutions can be regarded as self-matching.

Many-to-one matching can be defined as:
\begin{definition}
	\textit{
	A matching $\mu$ is a function from the set $V \cup P$ into the set of unordered families of elements of $V \cup P$ such that:
	\begin{enumerate}
		\item $|\mu(p)|=1$ for every person $p\in P$ and $\mu(p)=p$ if $p$ is unmatched;
		\item $|\mu(v)|=q_v$ for every institution $v\in V$; if the number of people $k$ in $\mu(v)$ is such that $k<q_v$, then $\mu(v)$ will have $q_v - k$ copies of $v$;
		\item $\mu(p) = v$ if and only if $p$ is an element of $\mu(v)$.
	\end{enumerate}}
\end{definition}
Here, $\mu(p_1)=v$ means that person $p_1$ is matched with institution $v$ and $\mu(v)=\{p_1, p_2, v, v\}$ means that the institution $v$ with the quota $q_v=4$ has been matched with two people $p_1$ and $p_2$ and has two unfilled matching positions. Pairwise stability in many-to-one matching is defined in the same way as one-to-one matching, and any coalition cannot block a stable matching.

\subsection{Many-to-Many}
A many-to-many matching problem refers to when the number of matchings for the agents on both sides are not restricted to one \cite{kong2019many}.

Suppose that two disjoint finite sets of agents are $S=\{s_1,s_2,...,s_n\}$ and $T=\{t_1, t_2,...,t_m\}$, with $q_s$ and $q_t$ being the respective quotas for agents $s\in S$ and $t\in T$. We generalise the definition of many-to-one matching as follows:

\begin{definition}
	\textit{
	A matching $\mu$ is a function from the set $S \cup T$ into the set of unordered families of elements of $S \cup T$ such that:
	\begin{enumerate}
		\item $|\mu(s)|=q_s$ for every agent $s\in S$; if the number of agents $k$ in $\mu(s)$ is such that $k<q_s$, then $\mu(s)$ will contain $q_s - k$ copies of $s$;
		\item $|\mu(t)|=q_t$ for every agent $t\in T$; if the number of agents $l$ in $\mu(t)$ is such that $l<q_t$, then $\mu(t)$ will contain $q_t - l$ copies of $t$;
		\item $s\in \mu(t)$ if and only if $t\in \mu(s)$.
	\end{enumerate}}
\end{definition}

In the real-world matching markets, there are some classical many-to-many cases, i.e., medical intern matching~\cite{sotomayor1990two} in the U.S., rider-driver matching in ride sharing~\cite{bin2019uroad}, and the teacher-student matching~\cite{kurino2020credibility}. Besides, pre-caching the various files for mobile users is also a many-to-many matching problem to improve the network performance in D2D netwrok~\cite{qian2020many}.

\subsection{Deferred Acceptance Algorithm}
\par  To achieve the goal of stable matching, The Deferred Acceptance (DA) algorithm was proposed and applied to the marriage markets and college admission problems~\cite{gale1962college}. Stable matching can also be achieved in many other markets by deploying the DA algorithm and its variants~\cite{islam2016stable,bendlin2019partners}.
\par
\begin{figure}[htb]
	\centering
	\includegraphics[width=0.45\textwidth]{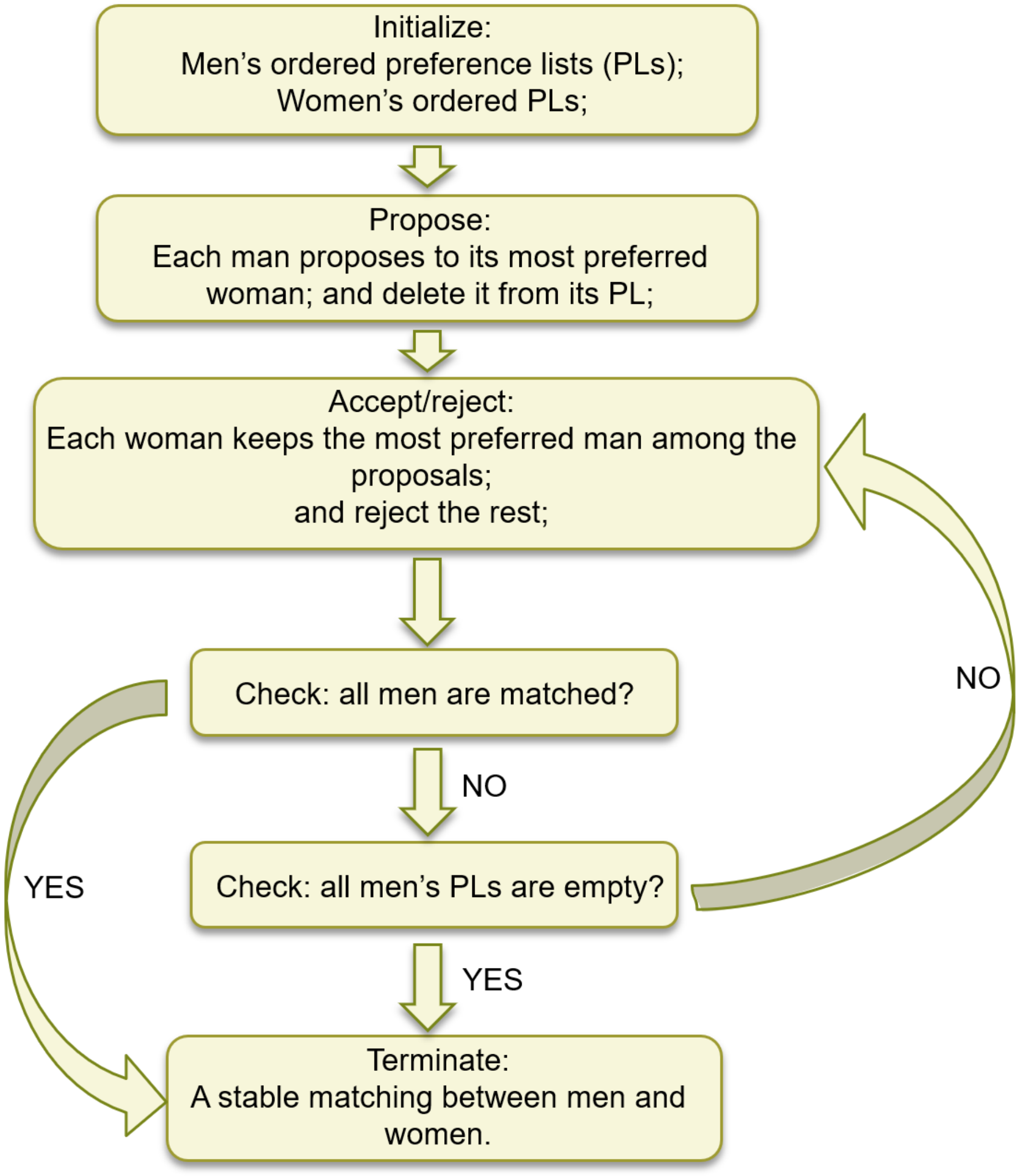}\\
	\caption{Flow chart of DA algorithm in marriage market.}\label{DA}
\end{figure}

 In the DA algorithm, the agents on one side propose a pair formation with the agents of the other side according to their preference, and an iterative procedure follows. The other set of agents reject the offers of any agents that are not acceptable to them, and each agent that receives more than one offer rejects all but its most preferred one. Any agent whose offer is not rejected at this point is tentatively matched with the agent they proposed to; however, their current request could be rejected during the next iteration if the agent on the other set receives a better offer. In the next iteration, any agent whose proposal was rejected at the previous step makes new proposals to the agents on the other side based on the preference list. An agent will continue to propose according to the preference list as long as there are acceptable agents on the other side that are single. The agents on the other side continue rejecting the old proposals if the more preferred one comes along. The DA algorithm stops when there are no rejected agents that can make new proposals, at which stage the whole matching process terminates, and all the tentative matchings become final ones. A flow chart of the DA algorithm in the marriage market is shown in Fig~\ref{DA}.
\par Due to its efficiency in solving conflicts of interest among agents in the market. Xu and Li~\cite{Xu2011seen} advocate using a stable matching framework instead of utility-based optimization to solve network problems. Besides, matching theory can help a network designer in selecting an appropriate matching model for a specific application in wireless communications~\cite{bayat2016matching}. With a general understanding of matching theory, many resource allocation problems can be regarded as matching problems between agents.

\section{Implicit Matching}~\label{sec3}
In this paper, we propose a new concept, implicit matching, which is defined as looking for a method to pair two or more objects together to have a similar appearance. Unlike explicit matching, implicit matching focuses on the process of calculating the matching score. This section will introduce some methods of implicit matching from the perspectives of retrieval matching, user-item matching, entity-relation matching, and image matching.
\subsection{Retrieval Matching}
\begin{figure}[h!]
	\centering
	\includegraphics[width=0.45\textwidth]{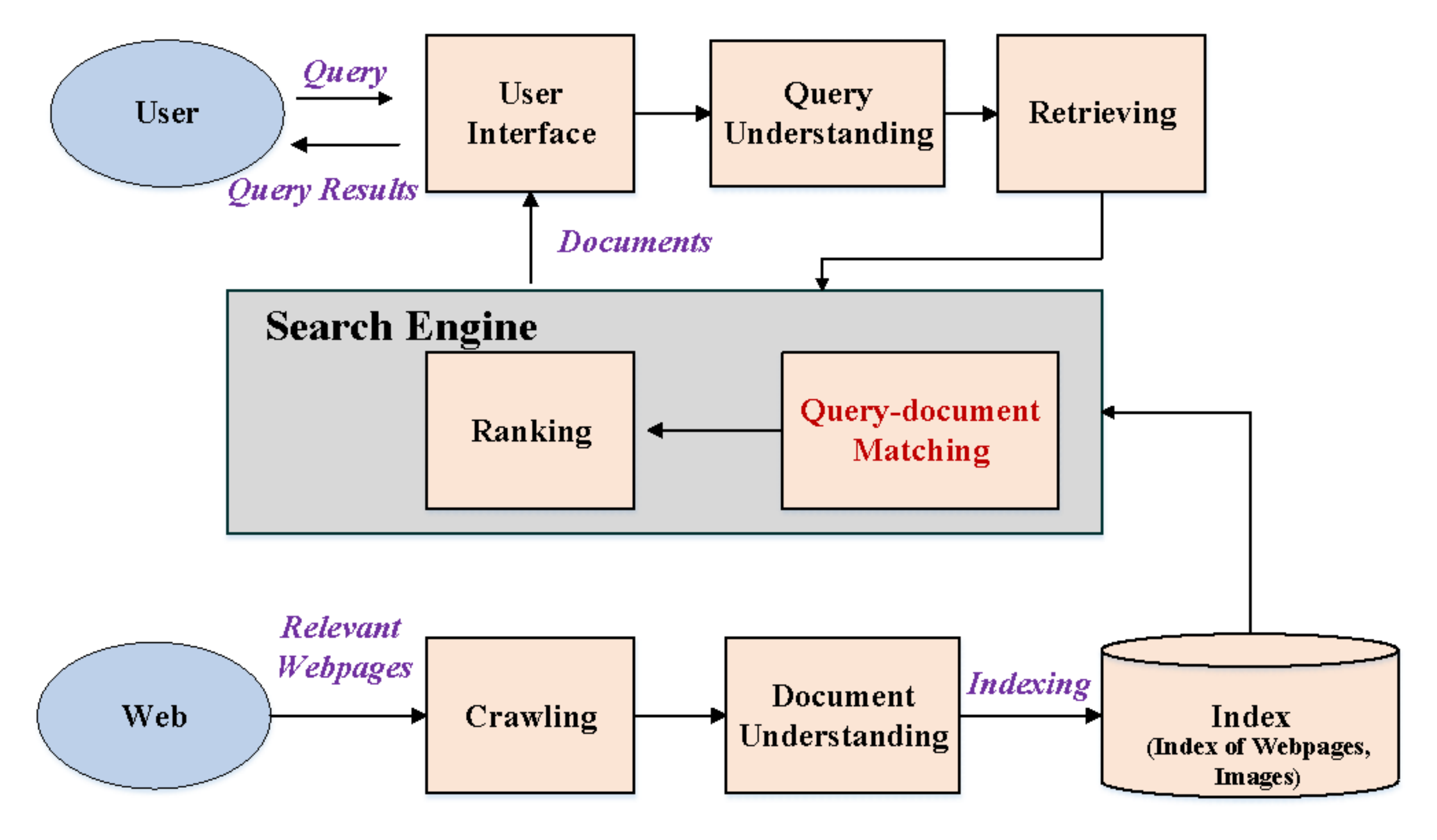}\\
	\caption{The process of information retrieval. The words with purple font means the transmitted data, and red font means the problem that need to be sovled.}\label{fig1}
\end{figure}

From the perspective of information retrieval, retrieval matching is how users actively input the query that expresses their needs clearly and obtain the information they want from the search engine's database. Fig.~\ref{fig1} describes the whole matching process of query and document in the search engine. Search engines act as a bridge between users and webpages. It matches the query searched by the user and the document in the database. The search engine extracts information from webpages and stores relevant information into the database. When the user enters the query, the search engine begins to analyze the query and finds the index database index that matches the query. The extracted webpages will be comprehensively ranked according to different conditions. Generally speaking, the search engine completes the final ranking through a series of steps, including query processing, query-document matching, relevance calculation, filtering, and adjustment, then returns the ranking results to users. In other words, information retrieval can be abstracted as a textual relevance matching problem between web pages and users' search queries. With the development of technology, the algorithms for textual relevance matching can be divided into three types: traditional, representation-based, and interaction-based matching algorithms. We will introduce them in the following subsections of the paper.

\subsubsection{Traditional matching algorithms}
Term Frequency-Inverse Document Frequency (TF-IDF) is a common weighting technique for textual relevance matching. It is a statistical procedure used to assess the importance of a word for a file set or one of the files in a corpus~\cite{ramos2003using}. The core idea of TF-IDF is to transform the problem of finding the matching degree of query $q$ and the document $d$ to finding the conditional probability that query $q$ comes from the document $d$. Given the individual document $d$ from the document corpus $D$ and the query $q$ containing the word $w$, TF-IDF calculates the importance (weight) of word $w$ in a document $d\in D$ as follows~\cite{salton1988term}:

\begin{equation}\label{eq1}
	TF-IDF_{w,d} = TF_{w,d} \cdot IDF_{w} = f_{w,d} \cdot \log\dfrac{|D|}{n(w)},
\end{equation}
where $f_{w,d}$ is the frequency with which $w$ appears in $d$ (TF component), $|D|$ is the number of documents in $D$, and $n(w)$ is the number of documents in $D$ that contain the word $w$ (IDF component). The value of TF-IDF is simply the product of the TF component and IDF component. Generally speaking, the larger the TF-IDF of a certain word $w$ in the document $d$, the higher its importance in this document. Therefore, by calculating the TF-IDF of each word in the document, the first few words with larger TF-IDF are considered as the keywords of this document. Variations of TF-IDF are often used in search engines to measure the degree of correlation between files and users' queries.

BM25 (Best Match 25)~\cite{salton1988term, robertson2009probabilistic} is a typical TF-IDF matching algorithm for evaluating the relevance between the query and documents based on probability. The main idea is to calculate the weights of the words in the query for each document and sum these weights over the set of words in the query to measure its relevance to the document. The relevance (matching) score $score (q,d)$ of a document $d$ in a query $q$ can be formulated as:

\begin{equation}
	score (q,d) = \sum_{w\in q} TF_{w,d} \cdot IDF_w.
\end{equation}
Here $TF_{w,d}$ is a normalised TF component, and $IDF_w$ is the weight for the word $w$ in the query. $TF(w,d)$ is given by:
\begin{equation}
	TF_{w,d} = \frac{f_{w,d}(k+1)}{f_{w,d}+k\left(1-b+b\dfrac{|d|}{avgl(d)}\right)},
\end{equation}
where $|d|$ is the length of a document $d$, $avgl(d)$ is the average length of the documents in the corpus that $d$ comes from, and $k$, $b$ are the parameters. The default $IDF(w)$ is generally calculated as:
\begin{equation}
	IDF_w = \log \frac{|D|-n(w)+0.5}{n(w)+0.5}.
\end{equation}
The meanings of $|D|$ and $n(w)$ are the same as Eq.~\ref{eq1}.

Latent semantic analysis (LSA)~\cite{dumais2004latent} is used initially in semantic retrieval to solve the problem of different words for the same objects and concepts (synonymy). In some cases, LSA is also called latent semantic indexing (LSI). The purpose of LSA is to discover the implied semantic dimension (can be understood as "topic" or "concept") from the text. The basic idea of LSA is to reduce high-dimensional documents to low-dimensional space, which is called latent semantic space. The steps of LSA can be summarized as:
\begin{enumerate}
	\item Analyze the document corpus and establish word-document matrix, of which rows are unique words and columns represent documents;
	\item Use a reduced-rank Singular Value Decomposition (SVD) to create a reduced-dimension approximation of the word-document matrix;
	\item Compute similarities of documents in the reduced dimensional space (latent semantic space).
\end{enumerate}
LSA can be used to compare different documents or match queries to the documents. The similarity of texts is semantic, so documents can be close even if they do not share the same words. The word-document matrix is usually created using the TF-IDF technique to assign weights to the words. The similarity of a document and a query represented as two vectors can be calculated using the cosine distance between them. Values close to 1 illustrate that they are very similar. Since LSA maps words and documents into latent semantic space, it can remove some noise in the original vector space and improve information retrieval accuracy.

Hofmann brings LSA into a probabilistic space and proposes Probabilistic Latent Semantic Analysis (PLSA)~\cite{hofmann1999probabilistic}, which uses the Expectation Maximization (EM) algorithm to estimate the parameters of the model. PLSA is based on the idea that the documents are modeled as multinomial probability distributions of topics, and the topics are modeled as multinomial distributions of words. The number of parameters in PLSA grows linearly with the number of training documents, making it prone to overfitting. The model cannot predict distributions of new documents as it learns the topic distributions only for the documents it has trained.

To overcome these problems, Beli et al.~\cite{blei2003latent} propose a three-level hierarchical Bayesian model named Latent Dirichlet Allocation (LDA). Being an extension of PLSA, LDA is also a probabilistic and generative model with documents modeled as mixture over a set of topics. As opposed to assuming that the topic weight parameters are fixed and unknown in PLSA, LDA treats them as a hidden random variable generated by a Dirichlet distribution.

\subsubsection{Representation-based algorithms}
Representation based information retrieval methods focus on learning the representations of the queries and documents. Sentences are encoded into their embedding without any cross-interaction~\cite{minaee2020deep}. An advantage of these methods is that they can be used for transfer learning to other natural language tasks. There are mainly two processes to complete the matching:
\begin{itemize}
	\item Calculate the representation $\phi(q)$, $\phi(d)$ for the query $q$ and the document $d$, respectively;
	\item Conduct the matching $M(\phi(q),\phi(d))$.
\end{itemize}

We will now introduce some typical methods for each process in detail. Huang et al.~\cite{huang2013learning} propose the Deep Structured Semantic Model (DSSM), which is a supervised learning method to learn the representations of queries and documents using a Deep Neural Network (DNN) framework. In the input layer of the model, the DNN processes the word vectors of queries or documents by taking advantage of word hashing to reduce the dimensionality of the vectors, after which it pushes the hashed features through multiple layers of non-linear projections. As a result, the queries and documents are mapped into concept vectors in the reduced-dimensional semantic space. The matching score between two vectors is measured using the cosine similarity of their respective concept vectors.

DSSM uses a bag-of-words approach to treat documents and queries, which ignores the word order and context information. Shen et al.~\cite{shen2014latent} propose a convolutional latent semantic model (CLSM, which can also be called CNN-DSSM) incorporating a convolution-pooling structure to generate the representations for queries and documents to overcome the latter drawback of DSSM. The major difference between CLSM and DSSM lies in the convolution and pooling layers of the neural network, through which CLSM can extract the contextual information for each word within a context window. However, it is challenging to preserve context information at long intervals due to the size limitation of the CLSM window (convolution core).

Palangi et al.~\cite{palangi2014semantic} suggest combining Long-Short-Term Memory (LSTM) with DSSM and propose LSTM-DSSM aiming at solving the disadvantage of CNN-DSSM's inability to capture long-range context features. It uses a variant of LSTM, LSTM with peephole~\cite{gers2000recurrent} to generate the error signal from the cosine similarity of the embedding vectors. Experiments on the information retrieval task demonstrate the effectiveness of the method.

It is acknowledged that recursive neural networks (RNNs) can embed text into fixed-length vectors, which have also shown good performance on various natural language processing tasks. Considering that RNNs require the input to be structured and make data preparation more complicated and time-consuming. Therefore, Choi et al.~\cite{ChoiYL18} devise a sentence encoder model to efficiently learn to compose task-specific tree structures from plain text data, whose architecture is a tree-structured long short-term memory (Gumble Tree-LSTM). This model introduces the composition query vector to compute the candidate parents' validity and selects the appropriate parent according to validity scores.

\subsubsection{Interaction-based algorithms}
Instead of directly learning the semantic representation vectors of queries and documents, interaction based methods let them interact in advance at the bottom layer and establish some basic matching signals, i.e., the matching of word and word, and then try to integrate these basic matching signals into a matching score. There are also two major steps to complete the matching:
\begin{itemize}
	\item Construct the basic low-level signals $s(q)$, $s(d)$ for the query $q$ and the document $d$, respectively;
	\item Aggregate the matching patterns $A(s(q),s(d))$.
\end{itemize}
Hu et al.~\cite{hu2014convolutional} adapt the convolutional strategy in matching sentences and propose two architectures ARC-I and ARC-II which can capture the structures of sentences at both the same and different levels. The first step for ARC-II is carrying out ARC-I, which aims to represent the sentences and compares the representations with a multi-layer perceptron (MLP)~\cite{bengio2009learning}. Specifically, it needs to set the window size and then extract the trigram vectors of two sentences. Firstly, by using the convolutional strategy, a matrix which is the matching signal of two sentences is constructed. By using continuous convolution and pooling, a vector of a certain length is produced. Finally, ARC-I uses MLP to provide the matching scores between sentences. However, as ARC-I keeps the interaction between two sentences until their final representations are created, it can lose important details for the matching. ARC-II overcomes this drawback by allowing the sentences to interact before their final representations are made.

Following~\cite{hu2014convolutional}, Pang et al.~\cite{pang2016text} propose MatchPyramid which is a convolutional model that tackles text matching problem through image recognition. The matching matrix which is obtained using cosine or dot-product similarity between word vectors is viewed as an image. MatchPyramid solves the problem that there is no matching signal for unigram in ARC-II.

Wan et al. \cite{wan2016match} propose to solve the text matching problem recursively. The interaction between two texts at any position is presented as a composition of interactions between their prefixes and the word level interaction at the position. To model the recursive structure of the matching, Match-SRNN method utilizing neural networks is introduced. It constructs a similarity tensor to capture word interactions and a spatial RNN with gated recurrent units is subsequently applied to it. Finally, the matching score is calculated based on the global interaction using a linear function. That is, given the matching score $h(\cdot)$ of the prefixes and the similarity score $s(\cdot)$ of the words, the matching score for the query and document of lengths $m$ and $n$ is calculated as:
\begin{equation}
	score(q,d)=Wh_{m,n}+b,
\end{equation}
where $W$ and $b$ are parameters and
\begin{equation}
	h_{i,j} = f(h_{i-1,j}, h_{i,j-1}, h_{i-1,j-1},s_{i,j}),
\end{equation}
where $i=1,...,m$, $j=1,...,n$.
In Match-SRNN, 2D-GRU~\cite{cho2014learning} is used for function $f$. The approach used in Match-SRNN approximates a dynamic programming process for information retrieval.

Parikh et al.~\cite{parikh2016decomposable} use attention mechanism to decompose the Natural Language Inference problem (NLI) into subproblems. The proposed model consists of three parts: attention, comparison, and aggregation. Assuming that the vector for each word in the query is $q= [a_{w_1},..., a_{w_m}]$, and the vector for each word of the document is $d = [b_{w_1},..., b_{w_n}]$, the elements of the query and the document are aligned by adopting the attention mechanism. By using a neural network function $f$, the aligned phrases $\{a_{w_i},B_{w_i}\}^m_{i=1}$ and $\{b_{w_j}, A_{w_j}\}^n_{j=1}$, where $B_{w_i}$ is the subphrase in $d$ that is aligned to $a_{w_i}$, and likewise for $A_{w_j}$, are compared and the comparison vectors $v_{1i}$ and $v_{2j}$ are produced. Summing up the comparison vectors, respectively, and applying another neural network function to the result, we can get the similarity score.

Wan et al.~\cite{wan2016deep} present an architecture to match two sentences with multiple positional sentence representations generated by a bidirectional long short term memory (Bi-LSTM). Through k-Max pooling and a multi-layer perceptron, the matching score is finally obtained by aggregating such interactions between these different positional sentence representations. The feature vector $q$ obtained by k-Max pooling is first fed into a full connection hidden layer to get a higher level representation $r$. Then, the matching score $s$ is calculated as:
\begin{equation}
r=f(W_{r}q+b_{r}), s=W_{s}r+b_{s},
\end{equation}

where $W_r$ and $W_s$ denote the parameter matrices, and $b_r$ and $b_s$ are corresponding biases.

In~\cite{yang2016anmm}, the authors propose the idea that semantic matching between a question and answer is mainly relevant with semantic similarity rather than spatial positions. Therefore, instead of position-shared weighting scheme in CNNs, they combine different matching signals and incorporate question term importance learning using attention mechanism with value-shared weighting scheme.

Although attention mechanism is helpful to capture the semantic relationship and properly align the elements of two sentences, simply using a summation operation in the attention mechanism cannot retain original features enough. Therefore, Kim et al. \cite{kim2019semantic} propose a densely-connected co-attentive recurrent neural network. The recurrent and co-attentive features are connected from the bottom to the top layer without any deformation. The proposed DRCN consists of three components: 1) word representation layer, 2)attentively connected RNN, and 3) interaction and prediction layer. In the interaction layer, the representations $p$ and $q$ for the two sentences $P$ and $Q$ are aggregated in various ways. Then, the final feature vector $v$ for semantic sentence matching is obtained as:
\begin{equation}
v=[p;q;p+q;p-q;|p-q|],
\end{equation}
wherein, all operations in the equation are performed element-wise to predict the relationship between two sentences.

Mitra et al.~\cite{mitra2017learning} first combine the interaction-based model and representation-based model. They use two separate deep neural networks, one of which matches the query and the document by using the local representation and the other using learned distributed representations. Experiments on the web page ranking task illustrate that the combination of the models outperforms the state-of-the-art methods.

In summary, most literal matching algorithms such as TF-IDF and BM25 mainly depend on the word coverage degree between two documents, which have several disadvantages, such as semantic limits, structural limits, and knowledge limits. Although LSA and other semantic analysis models can make up for some of the disadvantages, they cannot fully replace the literal matching models. With the successful application of deep learning in computer vision, speech recognition, and recommendation system in recent years, many researchers devote to applying deep semantic matching models to natural language processing tasks to reduce the cost of feature engineering. The representation-based model focuses more on constructing the representation layer, where the text is transformed into a unique overall representation vector. However, compared with the representation based methods that can calculate the document embedding in advance, interaction based methods cannot calculate the semantic vector of the text in advance during the online prediction task, which will result in high online computation costs.

\subsection{User-item Matching}
\begin{figure}[htb]
	\centering
	\includegraphics[width=0.45\textwidth]{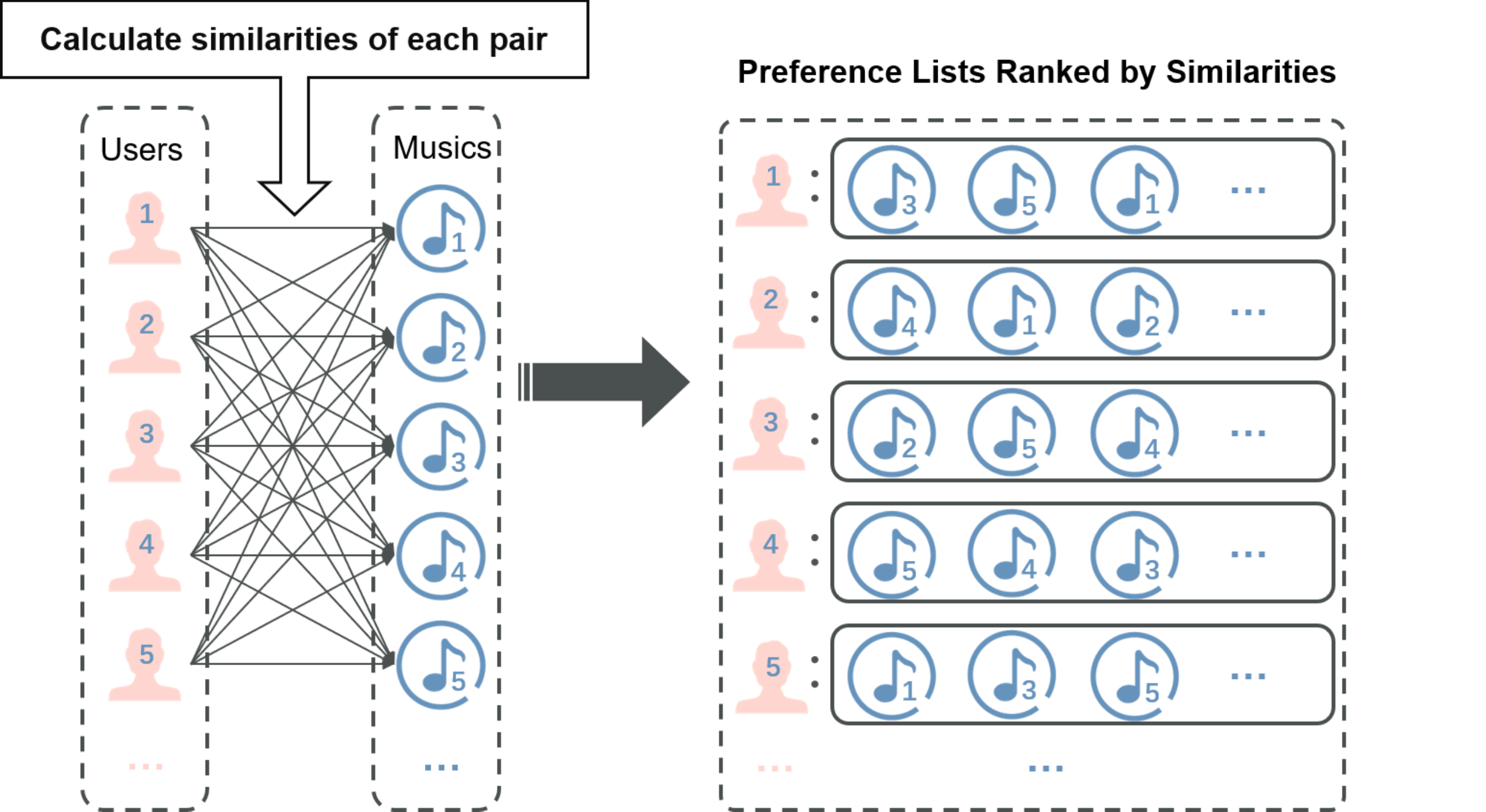}\\
	\caption{An illustration of recommending musics to users (Implicit Matching). First, the similarity (matching score) between each user and music is calculated according to the users' recordings, such as searching histories. Then, every user's preference list is obtained by ranking the similarity metric values, which can be used as a reference for recommending new music to users.}\label{implicitmatching}
\end{figure}

In the era of data explosion, a recommender system provides a convenient way for users to obtain their items of interest as accurately as possible~\cite{DLRS}. Therefore, personalized items recommendation are nowadays ubiquitous and have been performed in many practical applications, such as the recommendation of music in music player software, users on social websites and merchandise on shopping websites. Unlike searching, the recommendation is to push information or items to users by guessing their preferences or interests. An illustration of recommending music to users is shown in Fig.~\ref{implicitmatching}

\subsubsection{Basic algorithms}
\par Among all recommender systems, traditional recommender systems can be classified into three primary categories from the perspective of the data source~\cite{DLMRS}: collaborative filtering (CF), content-based recommendation (CR) and hybrid recommender system.
\par It is commonly acknowledged that CF is the most popular recommendation algorithm at present and is mainly classified from three perspectives, user-based, item-based, and model-based CF~\cite{Asocft}. The first two categories recommend items in terms of the similarity of users or items, while the model-based CF adopts some machine learning methods such as matrix factorization, clustering and restricted Boltzmann machine to build models. The emergence of content-based recommendation algorithms is early in its development and it can be performed in three steps.
\begin{itemize}
	\item Extract some features for each item (the content of the item) to represent it.
	\item Learn the user's preferable characteristics by utilizing the user's attitudes towards an item in the past.
	\item Recommend a list of the most relevant items for a user by comparing the user profile obtained in the previous step with the characteristics of the candidate items.
\end{itemize}
Matrix factorization~\cite{tran2018regularizing} is a standard method in recommender systems. A sparse user-item matrix is decomposed into a product of a user embedding matrix $U^{m\times k}=[p_1,...,p_m]^T$ and an item embedding matrix $V^{n\times k}=[q_1,...,q_n]^T$, where $m$ is the number of users, $n$ is the number of items, and $k$ is the number of latent factors, $k<min(m,n)$.  The predicted preference value of item $v$ by the user $u$, denoted by $\widehat{y}_{uv}$ is represented by the inner product of the user embedding $p_u \in \mathbb{R}^k$ and item embedding $q_v\in \mathbb{R}^k$,
\begin{equation}
	\widehat{y}_{uv}={p_{u}}^{T}q_{v}.
\end{equation}
We minimize the square of the difference between the predicted and the true preferences of the user, and to avoid the problem of over-fitting, a regularization function is added as a constraint. The loss function is therefore defined as follows:
\begin{equation}
	L=\underbrace{\sum_{u}\sum_{v}w_{uv}{(y_{uv}-\widehat{y}_{uv})}^{2}}_{Prediction~error} +\underbrace{\lambda(\sum_{u}{\parallel{p_u}\parallel}^{2}+\sum_{v}{\parallel{q_v}\parallel}^{2})}_{L2~regularizer}.
\end{equation}
Here $w_{uv}$ is a parameter that is used to balance the number of zero and non-zero values in the user-item matrix.

\subsubsection{Representation-based algorithms}
\par The methods of representation learning aim to learn the embeddings of the user and the item by the neural network. A simple inner product or the cosine value between their embeddings can be calculated as their final matching score. Methods of representation learning are introduced from two perspectives, CF methods where models are built merely based on the user-item interaction matrix, and methods of CF based on combining both user-item interaction and auxiliary information at the same time.

Sedhain et al.~\cite{Autorec} propose a novel autoencoder framework for collaborative filtering named AutoRec, which is a discriminative model based on autoencoders using a backpropagation algorithm to make the output of the model equal to the input. Compared with matrix factorization approaches, the item-based AutoRec model only embeds items into latent space. Subsequently, Wu et al. ~\cite{Cdaft} present a novel method called Collaborative Denoising Auto-Encoder (CDAE) for the top-N recommendation, which assumes that the observed user-item interactions are corrupted and, as such, is trained to reconstruct the data from the partially corrupted version of the preference set. CDAE differs from AutoRec by adding the userID to the input and excluding it from the reconstructed output layer, which is inspired by the thinking of SVD++~\cite{SVDpp}. To make the best of explicit ratings and non-preference feedback, Xue et al.~\cite{Dmfm} propose a new deep matrix factorization model with a neural network that projects users and items into low-dimensional vectors in the latent space. The input matrix for the model incorporates both explicit and implicit feedback from the users. A new loss function based on cross entropy is constructed that considers both explicit ratings and implicit feedback for the optimization.

The user $u_i$ and item $v_j$ are mapped to a latent space as follows:
	\begin{equation}
		p_i=f_{\theta_{n}^{U}}(...f_{\theta_{3}^{U}}(W_{U2}f_{\theta_{2}^{U}}(Y_{i*}W_{U1}))...),
	\end{equation}
	\begin{equation}
		q_j=f_{\theta_{n}^{I}}(...f_{\theta_{3}^{I}}(W_{V2}f_{\theta_{2}^{I}}(Y_{*j}^{T}W_{V1}))...).
	\end{equation}
Here $W_{U1}$ and $W_{V1}$ are the first layer weighting matrix for U and V, respectively, and $W_{U2}$ and $W_{V2}$ for the second layer, and so on. From the user-item interaction matrix $Y$, each user $p_i$ is represented as a high-dimensional vector of $Y_{i*}$, which represents the $i_{th}$ user’s ratings across all items. Each item $q_j$ is represented as a high-dimensional vector of $Y_{*j}$, which represents the $j_{th}$ item’s ratings across all users. The predicted $\widehat{Y}_{ij}$ is calculated from the dot product of $p_i$ and $q_j$:
\begin{equation}
\widehat{Y}_{ij} = p_{i}^{T}q_j.
\end{equation}

Additionally, a new loss function is designed to consider both explicit ratings and implicit feedback for better optimization, which is shown as follows:
\begin{equation}
		L=-\sum_{(i,j)\in Y^+\cup Y^-}(\frac{Y_{ij}}{max(R)}log\widehat{Y}_{ij}+ (1-\frac{Y_{ij}}{max(R)})log(1-\widehat{Y}_{ij})).
\end{equation}
Here, $R$ denotes the rating matrix, where $R_{ij}$ is the rating of user $i$ on item $j$. The max score in all ratings $max(R)$ is used for normalization, so that different values of $Y_{ij}$ have different influences to the loss.

In multimedia recommendation, to address the problem of implicit feedback, Chen et al.~\cite{Acf} propose a novel CF framework named Attentive Collaborative Filtering (ACF), which is an improvement based on SVD++~\cite{SVDpp}. By seamlessly incorporating two attention modules into neighborhood models, they can infer the underlying user preferences encoded in the implicit user feedback.

\subsubsection{Matching function-based algorithms}
The matching function learning methods are based on not directly learning the user and item's embeddings but by using a neural network to fuse the basic matching signal and subsequently obtain their matching score. Based on using CF models, He et al.~\cite{Ncf} can learn an arbitrary function from data and first devise a general framework named NCF for collaborative filtering based on neural networks. To design a better, dedicated interaction function for modeling the latent feature interactions between users and items, the offered solution is to let Generalized Matrix Factorization (GMF) and Multi-Layer Perceptron (MLP) share the same embedding layer, and then combine the outputs of their interaction functions. In order to provide more flexibility to the joint model, GMF and MLP are allowed to learn separate embeddings, and then be combined by concatenating their last hidden layer. The formulation of which is given as follows:
\begin{equation}
	\begin{aligned}
		\phi^{GMF}&=p_{u}^{G} \odot q_{i}^{G},\\
		\phi^{MLP}&=a_{L}(W_{L}^{T}(a_{L-1}(...a_{2}(W_{2}^{T}\left[\begin{matrix}
			p_{u}^{M}\\
			q_{i}^{M}
		\end{matrix}\right]+b_{2})...))+b_{L}),\\
		\widehat{y}_{ui}&=\sigma(h^{T}\left[\begin{matrix}
			\phi^{GMF}\\
			\phi^{MLP}
		\end{matrix}\right]).
	\end{aligned}
\end{equation}
Here, $\phi^{GMF}$ and $\phi^{MLP}$ are seperate embeddings learned by GMF and MLP, respectively. $\widehat{y}_{ui}$ is the final matching score by concatenating their last hidden layers. $p_{u}^{G}$ and $p_{u}^{M}$ denote the user embedding for GMF and MLP parts, and similar notations of $q_{i}^{G}$ and $q_{i}^{M}$  are used for item embeddings. $W_x$, $b_x$, and $a_x$ stand for the weight matrix, bias vector, and activation function respectively in the $x$-th layer perceptron of the MLP model. $h$ is the edge weight of the ouput layer.

Tay et al. \cite{Lrmlv} propose a new neural architecture named Latent Relational Metric Learning (LRML) for collaborative ranking with implicit feedback, which learns the relationship between users and items in metric space using latent relation vectors. The vectors are generated utilizing a newly devised auxiliary memory module - Latent Relational Attentive Memory (LRAM) controlled by the user-item interactions, thus making the generated vector specific to each user-item pair.

\par When using the feature-based models, it is crucial to capture interactions and relationships of the features. Considering that the linearity of a single Factorization Machine (FM) can be insufficient and the complex structure of deep neural networks may be difficult to train, He and Chua ~\cite{Nfmfs} propose a novel model named Neural Factorization Machine (NFM) for learning higher-order and non-linear feature interactions, which couples the utility of FM and neural network in modeling second-order and higher-order feature interactions respectively. Based on the novel proposed Bi-Interaction operation, this model can learn more informative feature interactions at the lower level. To discriminate the importance of different feature interactions, Xiao et al.~\cite{Afml} present a novel model named Attentional Factorization Machine (AFM), which uses the attention mechanism to learn the weight of feature interaction. In this work, the AFM improves both the representation ability and the interpretability of a FM model.

In conclusion, the basic methods mentioned above are mainly based on matrix factorization, while the last two kinds of algorithms are deep learning-based matching models. The difference between these models is as follows. Representation-based learning focuses on learning the representations of the two items to be matched. Specifically, in the process of recommendation, model structures such as AutoEncoder, MLP, CNN or RNN can be used depending on the available information like text, image, and voice. For matching function-based learning, the entities are matched at the bottom layer. The underlying matching signal is fused with a neural network to get the final matching score. In the recommendation process, the focus of this kind of algorithm is to find ways to combine features.

\subsection{Entity-Relation Matching}
Another important matching is entity-relation matching in the knowledge graph representation. Recent years have witnessed a proliferation of knowledge graphs in many real-world applications such as semantic parsing~\cite{semantic-parsing1, semantic-parsing2}, information extraction~\cite{information-extraction}, link prediction~\cite{xia2019random,wang2020model}, recommender systems~\cite{liu2015trucom,xia2016scientific}, question answering~\cite{question-answering1,question-answering2} etc. Knowledge graph (KG) representation aims at transforming the symbolized components (e.g., entity and relation in a triplet) to vector, matrix, or tensor, which is easy to manipulate by computer. We categorize the algorithms into three groups: factorization-based algorithms, neural networks-based algorithms and translational distance-based algorithms.
\par For a given triplet $\langle h, r, t \rangle$ that represents a piece of knowledge, where $h$, $r$ and $t$ denote the head entity, relation, and tail entity, respectively, a typical representation process first projects $h$, $t$ and $r$ into a continuous vector space based on some methods (randomly or another pre-processing process such as Word2vec~\cite{word2vec}). It then defines a matching function on this triplet to measure the similarity or plausibility between entities in the triplet. During the training process, by randomly replacing entities or relations, the final representation vectors in a "golden" triplet should achieve the max matching function compared with other "negative" training triplets. We find that from the perspective of matching, the knowledge graph representation process can be easy to understand. The reason behind this is that the essence of knowledge representation is to discover the relation among large amounts of information which is equivalent to matching.

\subsubsection{Factorization-based algorithms}
Based on the previous introduction, we can see that the basic components in knowledge graphs are entity and relation in a triplet. The matrix and tensor are two commonly used forms to represent a knowledge graph. Factorization-based models construct a distributed representation in low-dimensional space for each entity and each relation while preserving the relationship between them. Afterward, the specific tasks such as prediction can be performed by using appropriate similarity calculation formula like inner product.
\par  Sutskever et al.~\cite{BCTF} propose the Bayesian Clustered Tensor Factorization (BCTF) model which constructs the matching score according to the partition to which relation belongs. They assume entities in the same cluster should obey similar distributed representations. The matching degree of distributed representations should be determined by the value of $E^T_hWE_t$ where $W\in \mathbb{R}^{d \times d}$ is a weight matrix representing relation in a triplet, $E_h$ and $E_t$ are the vectors of head entity and tail entity.
\par RESCAL~\cite{RESCAL} model the entity and relation matching problem by a three-dimensional tensor. Each matrix slice can be factorized as a product of entity matrix and relation matrix. Unlike BCTF, RESCAL learns a unique space for entities in different domain which is independent of their occurrence in a relation. Furthermore, RESCAL can capture a more fine-grained matching between entities than BCTF.
\par In knowledge graph representation, we take each relation as a matching operator during the process of learning the embedding of entities and relations. However, there are multi-matching patterns in the real world. Like the matching theory in economics, the relation matching in knowledge graph can be one-to-one or many-to-many. BCTF and RESCAL use three-way models to construct the matching relations. However, the three-way model assigns each matching pattern the same capacity which can be friendly to frequent matching patterns and be problematic for rare matching patterns. The two-way model can be more feasible, in which each triplet $\langle h, r, t \rangle$ is decomposed into three binary interactions $(h,r)$, $(r,t)$, and $(h,t)$. To design a more general model for multi-matching patterns in knowledge graph, Alberto et al.~\cite{Tetac} propose Tatec (for Two And Three-way Embeddings Combination) that combines two-way model and three-way model for adopting different matching patterns in the knowledge graph representation. Their matching function consists of two parts: a Bigrams for modeling two-way matching and a Trigram for modeling three-way interactions. The matching function in Tatec is defined as
\begin{eqnarray} \label{Tetec}
	f(h,r,t) = B(h,r,t) + T(h,r,t),
\end{eqnarray}
where $B(h,r,t)$, $T(h,r,t)$ correspond to two-way and three-way matching, respectively.
\par Liu et al.~\cite{ANALOGY} propose another factorization-based model ANALOGY aiming at representing entity and relation through analogical inference. In an analogical inference task, one can find the correlation between two systems, and the unobserved part in one system can be inferred from the corresponding known part in another system. In ANALOGY, linear maps are used to represent the matching between entity and relation. The analogical structure is represented through the commutative properties of the linear maps. In a directed graph where the nodes are entities and the edges represent relations, the directed paths that share the same starting node and end node form the compositional equivalence. In addition to that, normal matrices are used for the linear maps for their convenient properties desirable for relation modeling. In ANALOGY, to reduce the matching problem's search space, the analogy is defined by adding normality and commutativity constraints to the matching function to capture the compositional equivalence.

\subsubsection{Neural networks-based algorithms} As previously mentioned, matching in knowledge graph representation is defined as mapping the entities and relations into a low-dimensional vector space, thereby obtaining the similarity score to measure how close the entities are. The factorization-based models process this problem practically in a linear way. However, entities and relations maintain considerable non-linear semantic information, which may not be represented by factorization. Therefore, some methods leverage neural networks to model the nonlinearity in knowledge representation.
\par SME (Semantic Matching Energy)~\cite{SME} transform the representation of entity and relation in knowledge graph to a semantic matching energy problem based on the energy-based learning theory~\cite{energy-based-learning}. According to this theory, the plausible triplets should be assigned low energies. The semantic energy matching function measures energy loss which transforms head entity to tail entity, therefore, the lower the value of the function, the more plausible the given triplet is. Triplets are first mapped to their embeddings $E_h, E_r, E_t$. The related-matching between entity and relation in a triplet can be formed as $g_1(E_h,E_r)$ and $g_2(E_t,E_r)$. The matching function defined in SME is:
\begin{equation} \label{SME-1}
	f(h,r,t) = g_1(E_h,E_r)^Tg_2(E_t,E_r).
\end{equation}
Here $g$ corresponds to a bilinear activation function defined as:
\begin{equation} \label{SME-2}
	g(x,y) = (W  \circ y^T)x^T + b,
\end{equation}
where $W\in \mathbb{R}^{p \times d \times d}$ is a weight tensor and $b \in \mathbb{R}^p$ is a bias.
\par Unlike SME, NTN~\cite{NTN} designs a novel neural network by changing the hidden layer to a bilinear tensor which aims at capturing the interactions between entities. Furthermore, instead of initializing the representation vectors by random sampling from a noise distribution, NTN chooses the average of word vectors to initialize the input. This improvement can preserve statistical characteristics of the input data. The matching function defined in NTN is:
\begin{equation} \label{NTN}
	f(h,r,t) = u_r^Tg(E_h^TW_r^{[1:k]}E_t+V_r[E_h,E_t]^T+b_r),
\end{equation}
where $g=tanh$ is an activation function and $W_r{[1:k]} \in \mathbb{R}^{d \times d \times k}$ is a tensor. The $E_h^TW_r^{[1:k]}E_t$ is a tensor product embedding in a vector space $\mathbb{R}^k$ with $i$th element computed as $E_h^TW_r^{[i]}E_t$. The other parameters $V_r\in \mathbb{R}^{k \times 2d} $, $u_r \in \mathbb{R}^{k} $ and $b_r\in \mathbb{R}^{k} $ are standard form of a neural network for relation $r$.

\par MLP (Multi Layer Perceptron)~\cite{MLP} represents entity and relation as a single vector by leveraging CNN. The weights in the neural network preserve the interactions between entities and the activation function capture the nonlinearity in the knowledge graph. The matching function in MLP can be defined as:
\begin{equation} \label{MLP}
	f(h,r,t) = \sigma(u^Tg(W^T[E_h,E_r,E_t])),
\end{equation}
where $W \in \mathbb{R}^{d \times 3d}$ represents the input layer weights and $u \in \mathbb{R}^{d}$ represents the hidden layer weights. $g$ is a non-linear function such as $tanh$. Compared with NTN, MLP model achieves almost the same performance but requires less parameters.
\par NAM (Neural Association Model)~\cite{NAM} uses a deep neural network to model the association between two events $A_1$ and $A_2$. This deep neural network with a multi-layer nonlinear activation function can be adopted to compute the likelihood that two events $A_1$ and $A_2$ are to be matched using conditional probability $P(A_2|A_1)$. For a triplet $\langle h, r, t\rangle$, event $A_2$ represents head entity $h$ and relation $r$, and $A_1$  is an event that shows whether the tail entity $t$ is true or false. For a binary classification task, the corresponding activation part can be a $sigmoid$ function, and for multiple output tasks, the activation part can be a $softmax$ function. Therefore, this architecture of a deep neural network can be general for a lot of specific tasks. There are two structures of NAM, one is the traditional DNN and the other is RMNN (Relation-Modulated Neural Network), which is more suitable for modeling multi-matching tasks in the knowledge graph representation.
\subsubsection{Translational distance-based algorithms} Inspired by word embedding work such as Word2Vec~\cite{word2vec}, translational distance-based models adopt the matching process to a translation operator from head entity to tail entity. An example for this intuition is that two triplets $\langle king, isA, man\rangle $ and $\langle queen, isA, woman\rangle $ with the same relation should infer the equation $king - man \approx queen - woman$.
\par Bordes et.al~\cite{TransE} propose TransE which assumes that the translation between entities can be seen as a natural mapping process in representing them. The matching score function in TransE is defined as:

\begin{equation} \label{TransE}
	f(h,r,t) = \|E_h+E_r-E_t\|_{L_2}.
\end{equation}
Here $E_h,E_r,E_t\in \mathbb{R}^d$, the distance measurement is computed by $L_2$ norm and the score function should be small if the relationship triplet $\langle h,r,t\rangle$ holds.
\par From the perspective of matching, we can find that the matching process of TransE can not deal with the reflexive matching, many-to-one and many-to-many matching. The reason behind it is that TransE ignores distributed representation of entities in different relations. To solve this, TransH~\cite{TransH} is proposed by projecting relation into a hyperplane space. In other words, compared with TransE which projects entity and relation in the same space, TransH maps the translation or matching operator to another space which distinguishes the different semantic roles one entity plays in different matching scenes. The head entity and tail entity are first projected into the relation-hyperplane as:
\begin{equation} \label{TransH1}
	E_{h\bot} = E_h - w_r^TE_hw_r, ~E_{t\bot} = E_t - w_r^TE_ttw_r,
\end{equation}
where $w_r$ is the normal vector to the hyperplane. Here $E_{h\bot}$ and $E_{t\bot}$ should be close in the projection hyperplane if they describe a plausible fact as shown in Fig.~\ref{Trans}. The score function in TransH is defined as:

\begin{equation} \label{TransH2}
	f(h,r,t) = \|E_{h\bot} +E_r-E_{t\bot}\|_{L_2}.
\end{equation}
Through another hyperplane, the matching process in TransH can be more fine-grained in representing the entity and relation in knowledge graph.
\par Although TransH constructs a hyperplane for relation, it still assumes the embeddings of entities and relations should be in the same space. However, entities and relations of knowledge graph preserve various semantic information in different scenarios such as $\langle apple,kindOf, fruit\rangle $ and $\langle apple,isA,company\rangle$. TransR~\cite{TransR} constructs different spaces for entity and relation. The matching process from entity space to relation space can be done by a specific projection matrix $M_r$, with the entity vector projections defined as follows:
\begin{equation} \label{TransR1}
	\hat{E}_h = E_hM_r, \hat{E}_t = E_tM_r.
\end{equation}
According to this, the matching function in TransR is defined as:

\begin{equation} \label{TransR2}
	f(h,r,t) = \|\hat{E}_h+E_r-\hat{E}_t\|_{L_2}.
\end{equation}
TransH and TranR focus their attention on modeling the diversity of relations in the semantic matching process which ignores the diversity of entities. In the matching theory we have introduced in other sections, the diversity of matching is related to the objects of matching. TransD~\cite{TransD} considers both the diversity of entity and relation. Two vectors are used to represent entity and relation in TransD, with one used to capture the semantic information and the other to construct the mapping matrix in translation.
\par From TransE to TransD, we find that these methods construct more fine-grained ways to model the complexity and diversity of entity and relation semantic information. However, the matching score functions defined in these methods are oversimplified by using inflexible metrics such as Euclidean distance. Such an over-simplified metric may miss a lot of information in modeling the nonlinearity with a spherical equipotential hyper-surface. TransA~\cite{TransA} defines a matching function using elliptical hyper-surfaces to better model complex embedding topologies of complex relations.  The matching function defined in TransA is:
\begin{equation} \label{TransA}
	f(h,r,t) = |E_h+E_r-E_t|^TW_r|E_h+E_r-E_t|,
\end{equation}
Where $W_r$ is a relation-specific symmetric weight matrix with non-negative elements. By distorting the original equipotential hyper-surfaces to an elliptical one, TransA enlarges the differences between entities at some dimensions, which can improve embedding complex entities in the knowledge graph.

This subsection mainly focuses on representing knowledge graph by mapping entities and relations into low-dimensional vectors while capturing their semantic meanings. Factorization-based algorithms process this problem essentially in a linear way. However, entities and relations maintain considerable non-linear semantic information, which may not be represented by factorization. Therefore, some methods leverage neural networks to model the nonlinearity in knowledge representation. In translational distance based models, the matching between entities and relations is regarded as a translation problem. It can be viewed as a semantic matching task in the natural language process.

\subsection{Image Matching}
Computer vision comes to its new age due to the dramatic boom of deep learning in recent years~\cite{zhou2020integrating}. Image matching as a basic algorithm can be applied to many research fields that require the ability to recognize and search for matching images, such as image retrieval, object tracking, face recognition, and object detection. Image matching is the process of making two images consistent in space so that the matching pixels in the two images are the same as the area to be matched. Here, we divide image matching algorithms into two types as~\cite{zitova2003image}: area-based matching and feature-based matching.
\subsubsection{Area-based algorithms}
Area-based methods are also called correlation-based methods or template matching. Area-based methods merge the feature detection step with the matching part, which is different from feature-based methods. Levine et al.~\cite{levine1973computer} incorporated an adaptive correlation window as a solution to object detection. They helped a robot to analyze the environment by representing three-dimensional objects in a scene into the depth map. Classical correlation measures were used within an adaptive window size. The area-based similarity metric can be used to define the most likely correspondence between the same sub-areas from two different views. In many applications, the dot product correlation function is used as a similarity metric, where the maximum score represents the best matching result.

Normalized Cross-Correlation(NCC) and its modifications are classical area-based methods. Gruen~\cite{gruen1985adaptive} proposed a powerful image matching technique that uses an adaptive least squares correlation. This technology can be applied to feature extraction, change detection, and line tracking of multi-spectral and multi-temporal images. Wu et al.~\cite{wu2019fast} proposed a fast, highly accurate NCC image matching algorithm. A constructed wavelet pyramid can reduce the searching and matching times of the feature point. An NCC image matching algorithm was then proposed to obtain the coarse matching points in the matching image.

NCC algorithm is considered a time-consuming approach. To increase computing speed, the Fourier~\cite{reddy1996fft} method is required for the images under different environments. In these images with noise, the Fourier method outperforms correlation methods. Fourier methods exploit the Fourier representation of images in the frequency domain.
A technology for quickly matching image with a number of images in a database, which extracts Fourier-Mellin phase features from images, was proposed by Ishiyama et al.~\cite{ishiyama2018fast} under the geometric changes of rotation and scale.

In the mutual information based registration method, the joint probability of comparable pixel intensities in the matching images is estimated. To detect and recognize the small dimensionality target, Yang et al.~\cite{yang2018image} proposed an image matching algorithm based on mutual information. The algorithm calculates the joint entropy of the matching image and the image to be matched. It then takes the coordinate with the maximum mutual information obtained in rough matching as the center position, and compares matching pixel by pixel to obtain the final matching score.

\subsubsection{Feature-based algorithms}
The work of image matching and image feature extraction can be traced back to 1981~\cite{moravec1981rover}. A corner detector was applied for stereo matching, but the Moravec detector was time-consuming and sensitive to noise. Soon after, Harris and Stephens~\cite{harris1988combined} developed a combined corner and edge detector based on the local auto-correlation function by improving the Moravec detector in 1988. The detector was composed of gradient information and eigenvalues of symmetric positive definite $2\times2$ matrix, and it was shown to perform with good consistency on natural imagery. Harris corner detector is sensitive to scale, which does not bring about a good matching performance for images of different sizes. Moreover, its application was limited to the stereo scene and short-range motion tracking.

Scale invariant feature transform (SIFT)~\cite{lowe2004distinctive} is one of the most widely used methods. It has good performance on matching and recognition due to its invariance to rotation, scale, and translation. Bay et al.~\cite{bay2008speeded} revealed that Speed-Up Robust Features (SURF) is an effective SIFT implementation method, which calculates the derivations of the image by applying integral images. Rublee et al.~\cite{rublee2011orb} proposed an efficient methods ORB alternative to SIFT or SURF, which is a very fast binary descriptor based on BRIEF. ORB is rotation invariant and robust to noise. Babri et al.~\cite{babri2016feature} made a comparative study on feature-based image matching algorithms. Furthermore, they concluded that the quality of features detected by SURF is better than SIFT because SIFT cannot match a large number of features it detects. For distorted images whose angle of rotation is proportional to 90 degrees, Karami et al.~\cite{karami2017image} showed that ORB and SURF outperform SIFT.

In previous years, research was mainly focused on face recognition under controlled conditions, among which simple classical methods provided excellent performance. Nowadays, the focus of research is on unconstrained conditions. Deep learning technology~\cite{guo2019survey} is becoming more and more popular because it provides strong robustness and can resist a large number of variations in the recognition process. In industry, face recognition as an application of image matching is generally divided into three steps: face detection, feature extraction, and feature matching. To extract high-quality features, face alignment usually comes before feature extraction. In the face matching part, two matching images are compared to obtain a similarity score, which gives the possibility that they belong to the same subject.

Facebook proposed Deepface~\cite{taigman2014deepface} in 2014. It is the foundation work of the deep convolutional neural network in the field of face recognition. The 3D model was used in the face alignment task. Then, the deep convolutional neural network implemented multi-class classification learning for the aligned face patch, using the classic cross-entropy loss function ($Softmax$) to optimize the problem. Finally, a fixed-length face feature vector was obtained through feature embedding. Later, Google proposed FaceNet~\cite{schroff2015facenet}, which used the Triplet Loss function instead of the Softmax for optimization on a hypersphere space to make the distance of clusters closer. Recently, ResNet~\cite{he2016deep} has become the most popular choice for many target recognition tasks. The main novelty of ResNet is the introduction of a building block that uses a shortcut connection to learn the residual mapping. ResNet facilitates the flow of information across layers, therefore the cross-layer connections allow training on deeper architectures. Nevertheless, face recognition is still facing many issues, such as racial bias in biometric that has not been thoroughly studied in deep face recognition~\cite{wang2019racial}.

In summary, the area-based methods are used to achieve dense matching without any obvious feature points detected from the image. They are more popular in highly overlapping image matching (such as medical image registration) and narrow baseline stereo images (such as binocular stereo image matching)~\cite{ma2020image}. Though the area-based methods have lower computational complexity, they limit the size of windows in large-scale images. Feature-based image matching can effectively solve the limitation problems in large viewpoint, wide baseline, and severe non-rigid image matching. It can be used for feature detection, discriminant description, and reliable matching, usually including transformation model estimation~\cite{ma2020image}. By combining multiple methods, more accurate and reliable matching solutions can be obtained. Therefore, compound deep learning approaches would be the solution to achieve more accurate results in some applications such as real-time detections~\cite{joglekar2012area}.

\section{Applications}~\label{sec4}
In previous sections, we introduced specific models and methods of two kinds of matching problems, which could be used to address a lot of serious problems in the real world. As a matter of fact, the matching theory is first proposed and studied in economics, which has path-breaking articles that developed intuitive algorithms. With the development of interdisciplinary research, the matching theory has been applied to more and more applications, and the term "matching" has extended its concepts to other fields. In this section, we present some practical applications in both explicit and implicit matching to better understand the concepts and methods in different matching problems.

\subsection{Applications in Explicit Matching}
\subsubsection{Matching in wireless networks}
Wireless networks consist of selfish and rational agents that naturally seek their maximum benefit from the system without considering other agents. In some complex wireless networks, various agents with different characteristics solicit communications with each other, where matching theory is particularly applicable to develop suitable and effective solutions. Here, we will introduce some of the significant applications of matching theory in wireless communication.

\textbf{Cognitive Radio Networks:}
Decentralized operation and efficient resource management are necessary for cognitive radio networks, which require licensed primary users to occupy channels that must be accessed by unlicensed secondary users. In other words, cognitive radio networks require stable solutions in matching licensed primary users and unlicensed secondary users. A number of recent works have corroborated the suitability of matching theory for cognitive radio~\cite{naparstek2014distributed,leshem2011multichannel}.

Leshem et al. \cite{leshem2011multichannel} took one of the first steps in applying matching theory into cognitive radio network. In this work, the association of licensed primary users with unlicensed secondary users is described as a one-to-one matching problem. The same utility function is used to get both sides' preferences. Hence, they find the stable allocation in a time-efficient way using a modified version of the DA algorithm. Later, Naparstek et al. \cite{naparstek2014distributed} extended this work from the perspective of energy efficiency.

\textbf{Device-to-Device (D2D) Communications:}
Device-to-Device (D2D) Communications is a technology devised to overcome the ever-increasing wireless capacity crunch. By introducing D2D in cellular networks, new challenges will arise in terms of interference management and resource allocation~\cite{gu2015matching}. Therefore, matching theory can be applied broadly in this area.

In \cite{gu2014cheating}, a form of "cheating" in the preference lists was incorporated to improve the DU's utilities. DUs can smartly change their preferences by cheating, thereby reaping more performance gains. In the final experimental results, the authors find that using such cheating strategies can simultaneously improve DU's and system utility compared with the DA algorithm.

\subsubsection{Matching in large firms }
Firms spend significant resources to hire the right employee and give different wages to workers in different occupations. Kelso and Crawford~\cite{kelso1982many} explore a general many-to-one matching model of firms for any number or type of workers. Kremer~\cite{kremer1993ring} explores O-ring production model which indicated several stylized facts, such as positive correlation among wages of workers in different occupations within a firm. Tervio \cite{tervio2008difference}, and Gabaix and Landier \cite{gabaix2008has} develop a matching model of firm size and CEO talent, and calibrate it using US data to analyze CEO pay. They show that the model exhibits a superstar property: small differences in talent can have a drastic impact on pay at the top.
\subsection{Applications in Implicit Matching}
\subsubsection{Expertise matching}

Within the context of big scholarly data~\cite{Bsdas}, expertise matching can be regarded as the process of finding the alignment between experts and queries. In other words, it is a process of finding individuals with the required knowledge and skills. Unlike information retrieval systems, expertise matching needs an expert retrieval system for facilitating knowledge exchange. Methodologically, existing methods can be divided into two classifications: probabilistic model and optimization model with multiple constraints, such as load balance, authority balance, and topic coverage~\cite{xia2014community}. Tang et al.~\cite{Emvco} regards the expertise matching as an optimization problem, the objective is to assign some experts to each query by satisfying certain constraints. Additionally, to validate their algorithms' effectiveness, they apply their framework to a practical conference system in terms of assigning experts to review papers. To support the rapid exchange of knowledge in innovation clusters, Babkin et al.~\cite{AMoOEMfFKE} propose a method of ontology-aided expertise matching based on a new methodology of ontology concepts matching. They also make several contributions to the advances of knowledge processing. Recently, Qian et al.~\cite{Wltmeioc} developed a model named weakly supervised factor graph (WeakFG) by considering two problems, how to trade off the degree between expertise and topic, and to what extent the invited users are willing to answer questions. They incorporate a number of correlations based on social identity theory into the WeakFG model, which combines expertise matching and correlations between experts. Furthermore, they design an online system to demonstrate the advantages of their proposed model.

\subsubsection{Question-answer Matching}
Baidu team proposes a multi-view question-answer model~\cite{zhou2016multi} in 2016, which is similar to the hierarchical structure and jointly considers the word-level matching and utterance-level matching. The SMN~\cite{wu2016sequential} model proposed by Microsoft Research Lab-Asia (MSRA) combines the representation based method and the interaction based method and integrates the word-level interaction and the segment-level interaction. Alibaba Group has officially launched the first exclusive AI customer service robot for Taobao and Tiancat apps users, AliMe Assis\footnote{https://consumerservice.taobao.com/online-help}. It is built on a multi-turn conversation model, MT-hCNN~\cite{qiu2018transfer} using convolutional neural networks. Zhang et al.~\cite{zhang2018modeling} focus on retrieval-based response matching for multi-turn conversation and propose the DUA model. It transmits important information in each utterance by introducing self-matching attention. After matching the response with the refined utterance and sharp turns aggregation, the final matching score can be obtained. Zhou et al.~\cite{zhou2018multi} from the Baidu research team propose Deep Attention Matching Network (DAM) model to realize response selection of chat robot in the multi-turn conversation understanding. It uses a transformer encoder to get the multi-granularity representation of the text. Then, it designs two interactive ways to get self-attention-match and cross-attention-match alignment matrices.
\subsubsection{Recommender Systems}

Recommender systems, especially deep recommender systems, have been widely used in industry and e-commerce scenarios~\cite{liu2018artificial}. An efficient recommender algorithm is at the core of big businesses for advertisements, media services, and online retailers, promotes business growth, and brings many economic benefits. The following introduces some typical recommender systems at the industry level.

Google proposes DCN (Deep\&Cross Network) to predict user click through rate (CTR). Taking advantage of DNN models, DCN introduces a novel cross network to learn bounded-degree feature interactions. DCN contains four critical parts, including embedding and stacking layer, cross network and deep network, and combination output layer. Feature crossing at each layer makes DCN require no manual feature engineering. Similarly, Facebook adopts DLRM (Deep Learning Recommendation Model)~\cite{naumov2019deep} for a personalized recommendation. DLRM combines collaborative filtering and predictive analysis. Specifically, DLRM first encodes user features as one-hot vectors, and obtains user representations by embedding lookup. The initial feature embeddings will learn the distinguishing feature representations in the process of model optimization. Finally, feature representations and their interactions are inputted in MLP, and the final click probability is predicted by the sigmoid function. Other typical algorithms for CTR, including LS-PLM~\cite{gai2017learning}, DeepFM~\cite{guo2017deepfm}, and NFM~\cite{he2017neural} have been applied to different advertising scenes such as Alibaba.

From the perspective of media services, personalized recommendation plays an important role in social media. For example, 80\% of the content that users watch on Netflix comes from recommendations. Netflix's recommendation system is divided into three parts: offline, nearline, and online. From offline to online, the real-time performance of data increases, while the scale and processing capacity of data decreases. Several core recommendation algorithms used by Netflix include Personalized Video Ranker (PVR), Top-N Video Ranker, Trending Now, Continue Watching, and Video-Video Similarity\footnote{https://netflixtechblog.com/}. YouTube is one of the largest and most complex industrial recommendation systems. It is designed to help more than 1 billion users discover personalized video content from a growing collection of videos. Before deep learning, YouTube's recommendation system was mostly based on user profiling and collaborative filtering~\cite{baluja2008video,davidson2010youtube,bendersky2014up}. The core deep learning algorithm~\cite{2016Deep} of the system consists of two neural networks, i.e., candidate set generation and deep ranking. The deep collaborative filtering model can effectively assimilate features and model their interactions, solving large scale problems, freshness, and noise in YouTube recommendation.

\section{Future Trends and Challenges}~\label{sec5}
In this section, we make a detailed discussion about the future trends and challenges of matching algorithms based on two key matching factors.
\subsection{Preference list}
The preference list is hidden in large-scale data with the changes and development of matching. Therefore, how to extract or infer the preference list which can reflects the real expectation rank of agents is crucial in existing matching scenarios.

\subsubsection{Information fusion for preference list inference} 

Instead of giving a strict or clear rank order of preference for other agents in matching, the preference list is hidden in various data sources. Many matching algorithms are devoted to inferring a list that can reflect the real preference of agents. In the future, various sources of data will be collected or fused in matching systems and matching scenarios with the continuous deepening of computational intelligence. Therefore, it is worth exploring how to effectively fuse these heterogeneous data to obtain the preference list. A number of works have been proposed to explore this trend in retrieval matching~\cite{infofushion-1-1, infofushion-1-2, infofushion-1-3}, user-item matching~\cite{infofushion-2-1, infofushion-2-2, infofushion-2-3}, entity-relation matching~\cite{infofushion-3-1, infofushion-3-2}, and image matching~\cite{vo2019unsupervised,manzo2019graph}. However, extracting the decisive features from the problematic data such as data with missing values, noise, or outliers to complete the task of preference list inference poses a great challenge~\cite{Challenge-1}.

\subsubsection{Dynamic preference list inference} Another problem to be solved in preference list inference is its uncertainty. On one hand, the real preference information of matching agents may be hidden in various sources of data. On the other hand, this preference list may change with time. The typical instance in user-item matching is that users may change their preferences for items after sales promotion~\cite{dynamic-1}. Therefore, interactive matching systems such as interactive recommendation~\cite{dynamic-2}~\cite{dynamic-3}~\cite{dynamic-4}, interactive information retrieval~\cite{dynamic-5}~\cite{dynamic-6} have been witnessing a proliferation of attention in recent years to collect the feedback of matching agents to infer a more accurate preference list. Therefore, the challenge behind this task is to design an effective framework that can collect feedback information (e.g., incentive mechanism~\cite{dynamic-7}) and feedback information analysis (e.g., implicit feedback~\cite{dynamic-8}).

\subsection{Matching principles}
The matching principle can be understood as the expected outcome of the matching problems. For example, stability is the basic principle when researchers design matching algorithms for men and women in marriage markets. While in implicit matching, the ultimate goal of the model is to match the most similar pairs.
In a nutshell, a matching algorithm develops in accordance with application scenarios. Compared with classical matching market theory, data accessibility leads to more loose and complicated matching methods. Here, we outline three existing challenges when designing matching algorithms, namely fairness, interpretability and privacy-protection.
\subsubsection{Fairness}
Fairness is the inherent topic in matching. In classical matching theory, it has been proved that there is not a solution to guarantee fairness for both sides under a minimal set of axioms~\cite{masarani1989existence}. However, with the widespread use of AI systems in our daily life, it is important to consider fairness while designing matching algorithms, especially algorithms based on deep learning. The bias of an unfair algorithm usually comes from the heterogeneous data~\cite{fairness-1}. A matching algorithm trained on biased data may lead to unfair and inaccurate results. The fairness principle of matching algorithms will attract increasing attention in the future~\cite{fairness-2, fairness-3, fairness-4}. In addition, a clear definition of fairness and bias, as well as the dataset of unfairness samples still need to be explored in specific matching problems~\cite{fairness-1}.
\subsubsection{Interpretability}
Interpretability is also an important consideration when designing matching algorithms, which can also be understood as the rationalizability in AI system~\cite{ong2019air}. The classical matching market theory is explainable inherently, as it is designed based on the known preference lists. However, most implicit matching algorithms based on deep learning are black-box, which are incapable of answering 'how' the preference lists are inferred. Therefore, the interpretability of existing matching algorithms should be explored to design more credible matching algorithms. However, the heterogeneity of data in existing matching systems poses great challenges for implicit matching algorithms when considering interpretability~\cite{Inte-1}. Some preliminary related works about interpretable matching algorithms can be found in~\cite{Inte-2, Inte-3, Inte-4}.
\subsubsection{Privacy-protection}
As we have discussed previously in this section, it is difficult to obtain accurate and personalized preference lists with an increasing amount of information incorporated into matching systems. For example, in POI (Point of Interest) recommendation system, the location information, contact list information, and check-in information can be accessed and collected only if the platform gets the permission of users. Otherwise, it may cause the problem of privacy information disclosure. Therefore, protection of the privacy information is an important principle that need to be considered when designing implicit matching algorithms~\cite{privacy-1, privacy-2, privacy-3, privacy-4}. However, in the matching systems, balancing the cost in protecting the privacy information and benefits is still a controversial topic.

\section{Conclusions}~\label{sec6}
In this survey, we systematically summarized the common real-world matching problems and divided the matching problems into two categories according to the availability of the preference lists: namely explicit matching and implicit matching. In explicit matching, the matching problems are classified according to the agent requirements, namely one-to-one, many-to-one and many-to-many. In implicit matching, we mainly presented some common matching problems such as retrieval matching, user-item matching, entity-relation matching, and image matching. To better understand the concepts and methods, we introduced some practical applications for both categories of matching problems. Additionally, the future trends and challenges are also discussed according to two key matching factors. This article is expected to provide a comprehensive overview of matching problems and models suitable for the demands in various practical scenarios to address technical challenges in current and future matching problems.

\section*{Acknowledgement}
The authors would like to thank Jianshuo Xu and Jiaxing Li for their help with the first draft of this paper.

\ifCLASSOPTIONcaptionsoff
  \newpage
\fi



\bibliographystyle{IEEEtran}

\bibliography{BIB}

%
%
%

%
\begin{IEEEbiography}[{\includegraphics[width=1in,height=1.25in,clip,keepaspectratio]{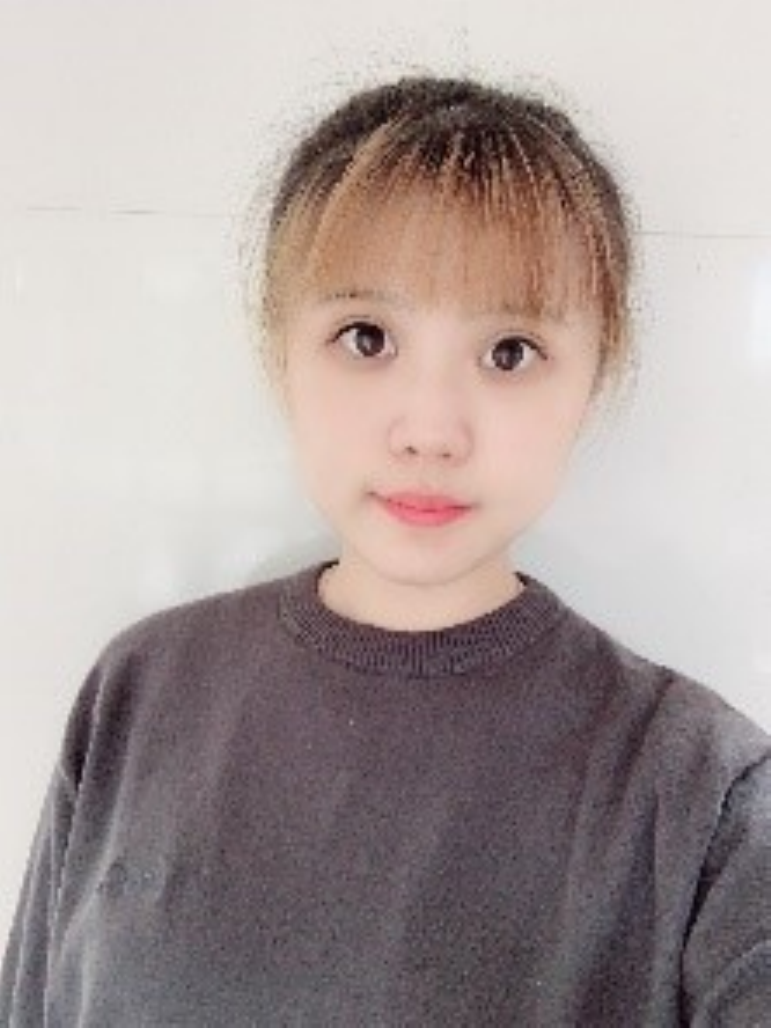}}]{Jing Ren}
%
received the Bachelor’s degree from Huaqiao University, China, in 2018, and the Master degree from Dalian University of Technology, China, in 2020. She is currently pursuing the Ph.D. degree in School of Engineering, IT and Physical Sciences, Federation University Australia. Her research interests include data science, social computing, and graph learning.
\end{IEEEbiography}
\begin{IEEEbiography}[{\includegraphics[width=1in,height=1.25in,clip,keepaspectratio]{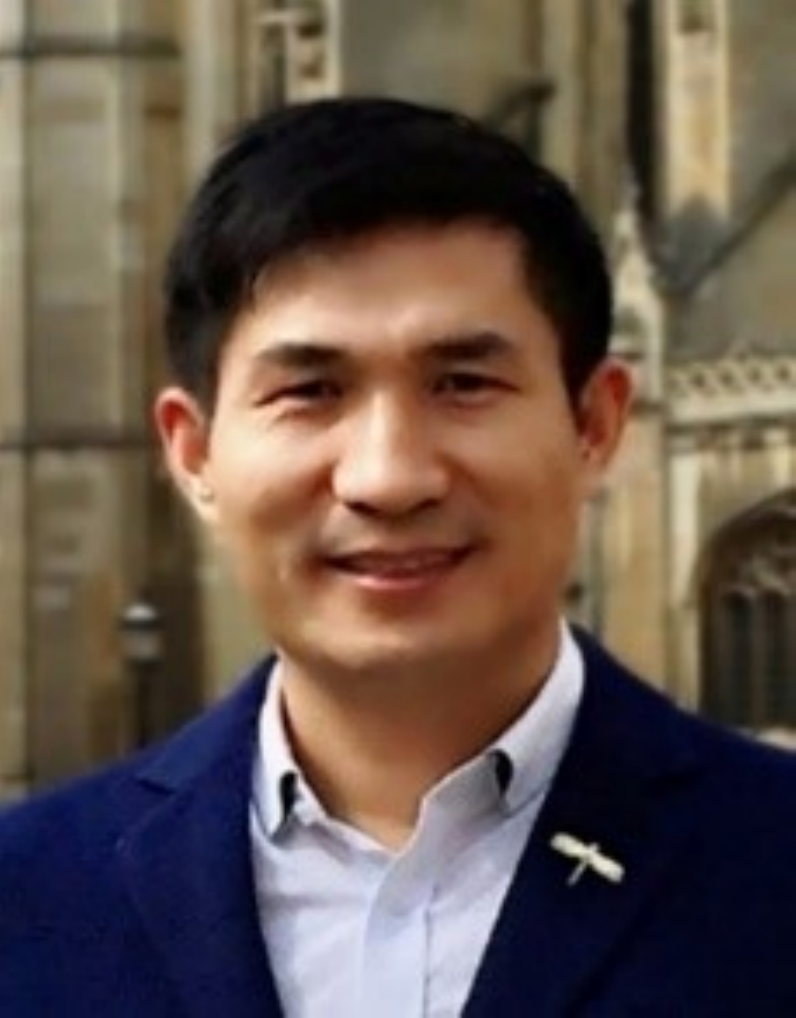}}]{Feng Xia}
	%
(M'07-SM'12) received the BSc and Ph.D. degrees from Zhejiang University, Hangzhou, China. He is currently an Associate Professor and Discipline Leader in School of Engineering, IT and Physical Sciences, Federation University Australia. Dr. Xia has published 2 books and over 300 scientific papers in international journals and conferences. His research interests include data science, social computing, and systems engineering. He is a Senior Member of IEEE and ACM.
\end{IEEEbiography}
\begin{IEEEbiography}[{\includegraphics[width=1in,height=1.25in,clip,keepaspectratio]{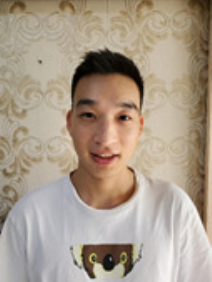}}]{Xiangtai Chen}
	%
received the B.Sc. degree in educational technology from Northwest Minzu University, Lanzhou, China. Currently, he is pursuing a master’s degree in software engineering at Dalian University of Technology, Dalian, China. His research interests include big scholarly data, data mining, and analysis of complex networks.
\end{IEEEbiography}
\begin{IEEEbiography}[{\includegraphics[width=1in,height=1.25in,clip,keepaspectratio]{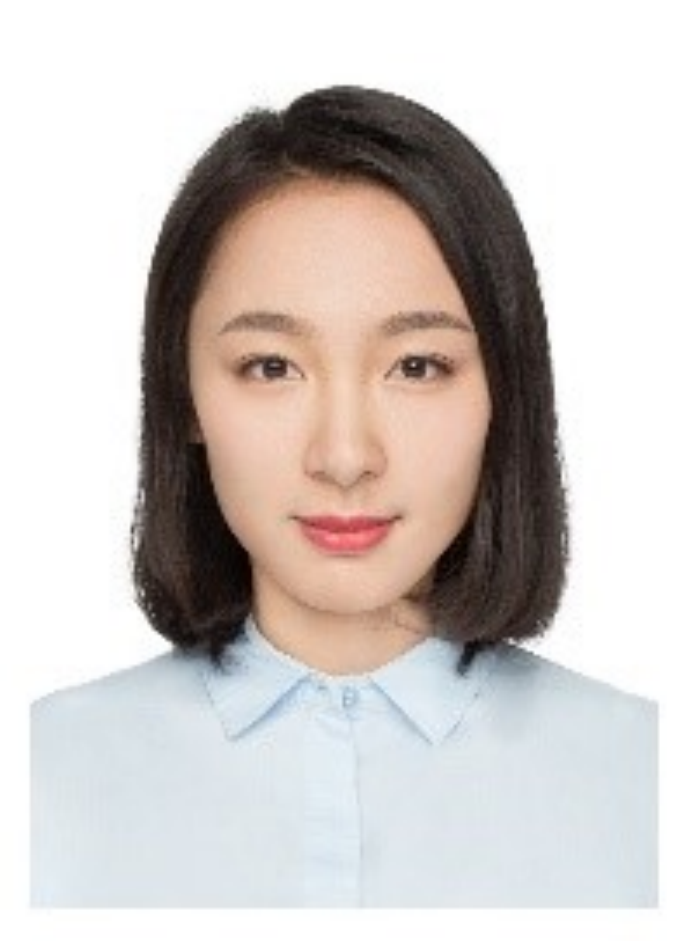}}]{Jiaying Liu}
	%
	received the BSc degree in software engineering from Dalian University of Technology, China, in 2016. She is currently working toward the Ph.D. degree in the School of Software, Dalian University of Technology, China. Her research interests include data science, big scholarly data, and social network analysis.
\end{IEEEbiography}
\begin{IEEEbiography}[{\includegraphics[width=1in,height=1.25in,clip,keepaspectratio]{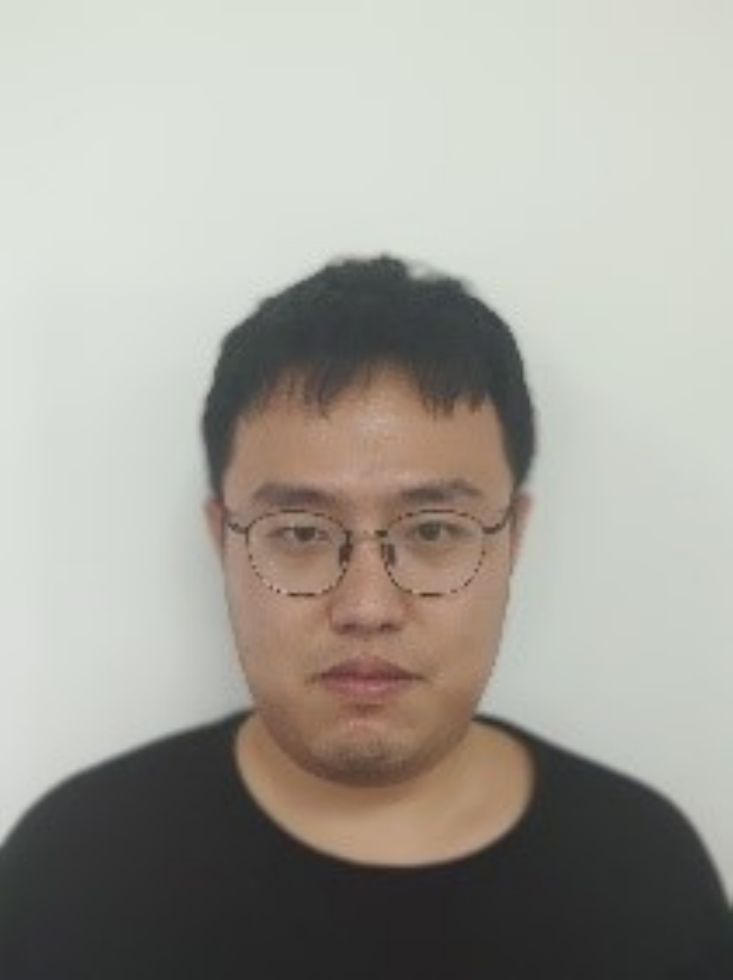}}]{Mingliang Hou}
	%
	 received the B.Sc. degree from Dezhou University and the M.Sc. degree from Shandong University, Shandong, China. He is currently pursuing the Ph.D. degree in software engineering with the Dalian University of Technology, Dalian, China. His research interests include network science, data science, and urban computing.
\end{IEEEbiography}
\begin{IEEEbiography}[{\includegraphics[width=1in,height=1.25in,clip,keepaspectratio]{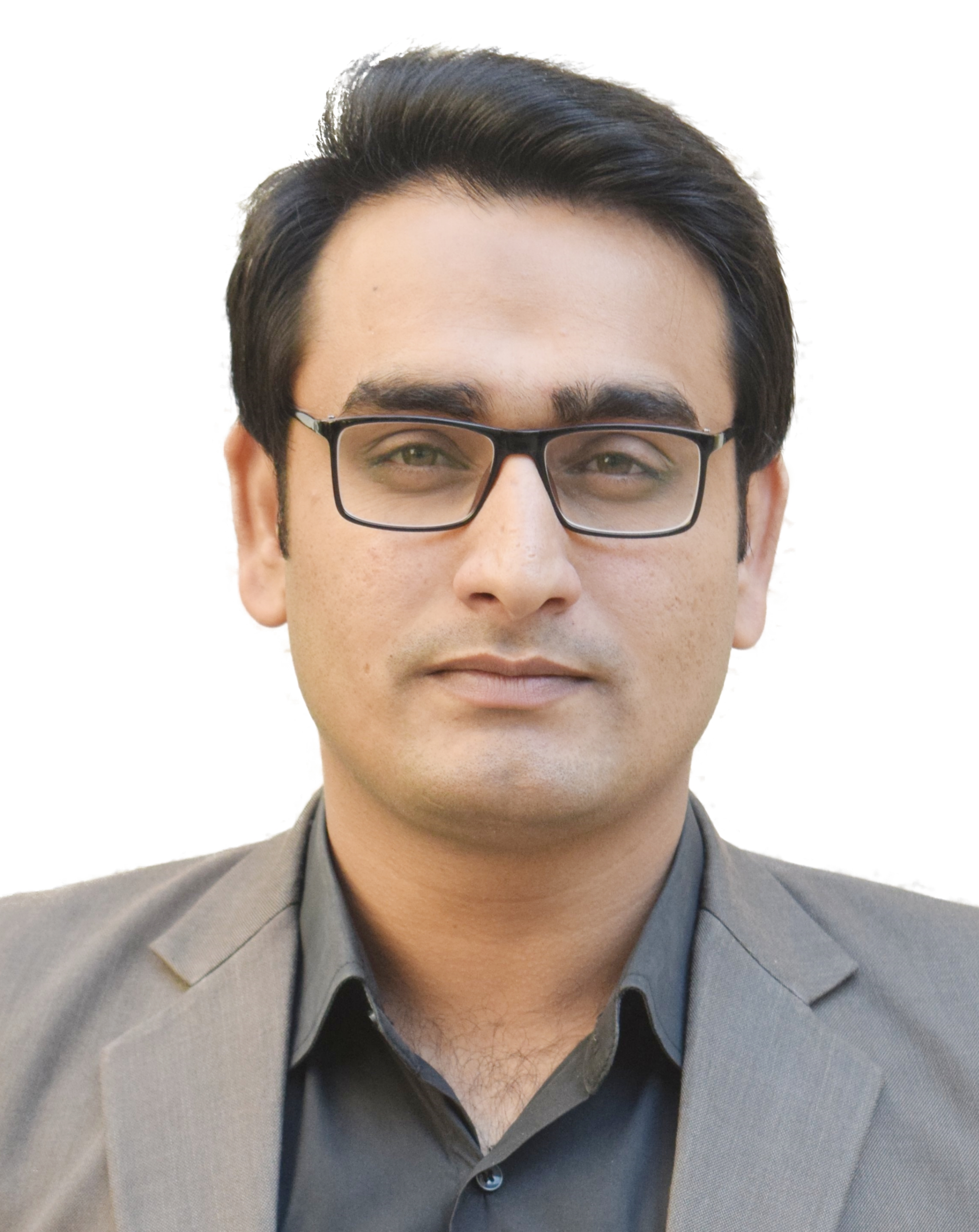}}]{Ahsan Shehzad}
	%
	was born in 1990. He received the B.S. degree in computer science systems from Bahauddin Zakariya University, Multan, Pakistan, in 2015, and the M.S. degree in computer science from Air University Islamabad, Pakistan, in 2018. He is currently pursuing the Ph.D. degree in software engineering with the School of Software, Dalian University of Technology, China. His research interests include digital image and video analysis, object re-identification, machine learning, and data science.
\end{IEEEbiography}
\begin{IEEEbiography}[{\includegraphics[width=1in,height=1.25in,clip,keepaspectratio]{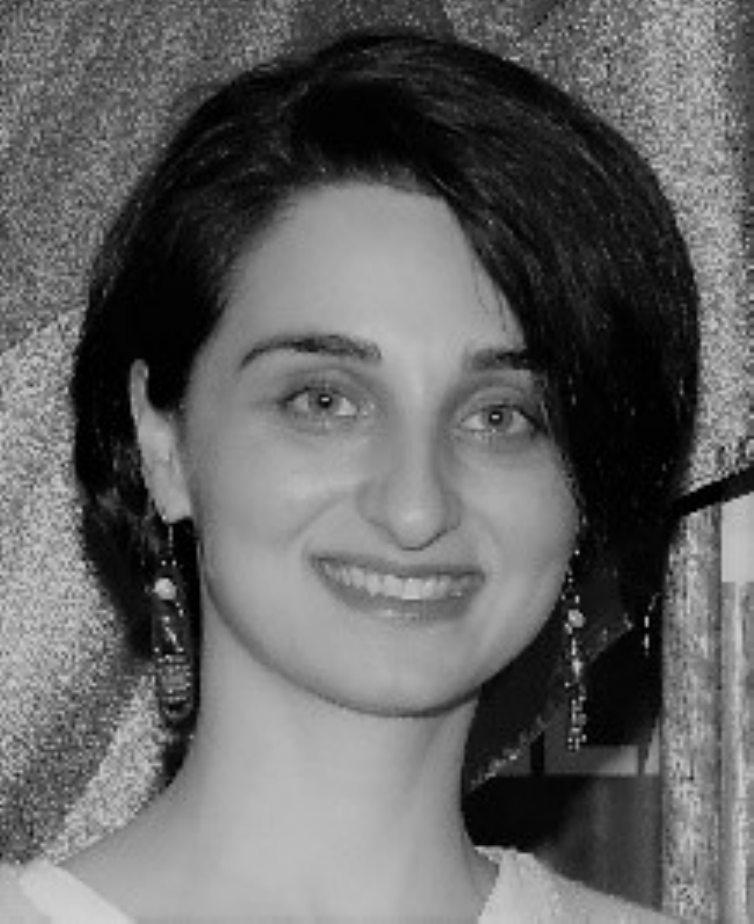}}]{Nargiz Sultanova}
	%
	is a Lecturer in Mathematics at the School of Engineering, IT and Physical Sciences, Federation University Australia. Nargiz received a Bachelor degree in Applied Mathematics from Baku State University in Baku, Azerbaijan, a Master degree in Mathematics and a PhD degree in Optimization from the University of Ballarat (currently Federation University Australia). Her main research interests lie in the area of nonsmooth optimization and its various applications. Nargiz is a member of AustMS and its ANZIAM division.
\end{IEEEbiography}
\begin{IEEEbiography}[{\includegraphics[width=1in,height=1.25in,clip,keepaspectratio]{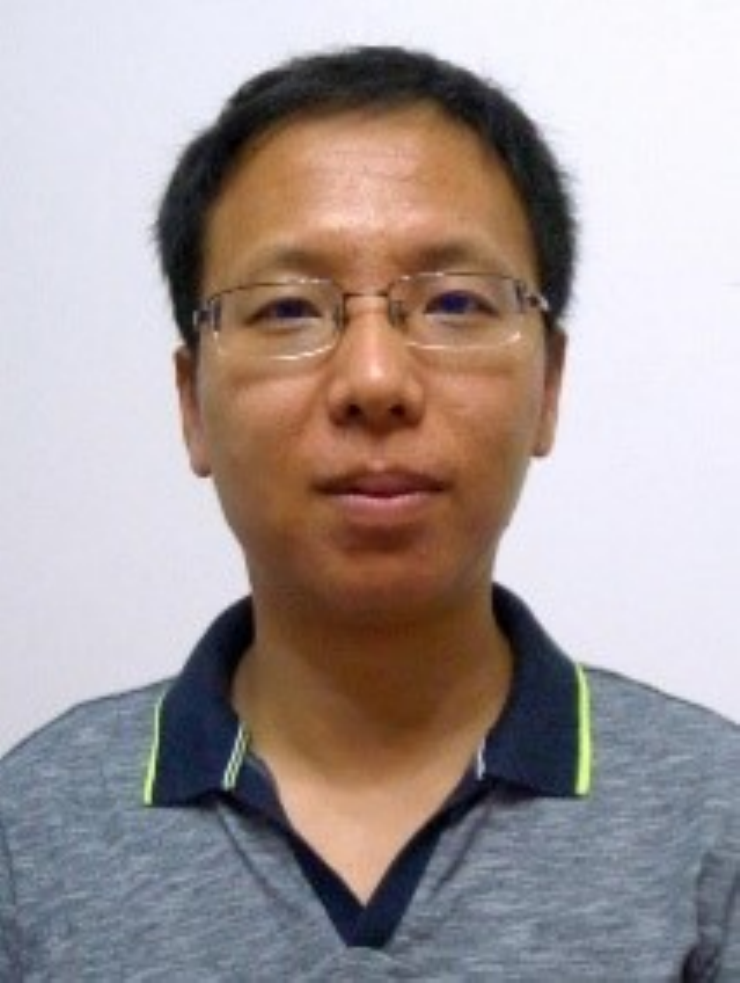}}]{Xiangjie Kong}
	%
	(M'13-SM'17) received the B.Sc. and Ph.D. degrees from Zhejiang University, Hangzhou, China. He is currently a Professor with the college of computer science and technology, Zhejiang University of Technology, China. He has published over 100 scientific papers in international journals and conferences. His research interests include network science, data science, and computational social science. He is a Senior Member of the IEEE and CCF and a member of ACM.
\end{IEEEbiography}
%

%




\end{document}